# High mechanical strength Si anode synthesis with interlayer bonded expanded graphite structure for lithium-ion batteries

Wenhui Lai [a], Jong Hak Lee [b], Lu Shi [a], Yuqing Liu [c], Yanhui Pu [a], Yong Kang Ong [b], Carlos Limpo [a], Ting Xiong [c], Yifan Rao [a], Chorng Haur Sow [b,c], Barbaros Özyilmaz [a,b,c,d,*]

[a] *Department of Material Science & Engineering, National University of Singapore, Singapore 117575, Singapore*
[b] *Center for Advanced 2D Materials, National University of Singapore, Singapore 117546, Singapore*
[c] *Department of Physics, National University of Singapore, Singapore 117551, Singapore*
[d] *Institute for Functional Intelligent Materials, National University of Singapore, Singapore 117544, Singapore*

* Corresponding author.
   *E-mail address:* barbaros@nus.edu.sg (B. Özyilmaz).

**Abstract:** On the verge of commercial success for silicon-based batteries, the development of silicon-based anodes has been largely impeded by the sluggish synthesis strategies (e.g., small tap density, low yield efficiency, and inadequate mechanical strength), and poor electrochemical performance (e.g., low electrode loading, insufficient areal capacity, and weak cycling stability) due to the extreme volume expansion. Here, we present an enhanced spark plasma sintering technology for the synthesis of expanded graphite and silicon nanoparticles composite monolithic blocks with interlayer welding. Such a highly scalable process enables a high-tap density electrode material of 1.68 g cm$^{-3}$ (secondary clusters: 1.12 g cm$^{-3}$) and a production yield of 87 g per 30 min process cycle. Notably, the interlayer welding structure offers not only vigorous ion and electron transport, but also robust mechanical stability (Vickers hardness: 658 MPa, Young's modulus: 11.6 GPa) to regulate and accommodate silicon expansion, thus achieving a superior areal capacity of 2.9 mAh cm$^{-2}$ (736 mAh g$^{-1}$) and a steady cycle life (93% after 100 cycles) under an active mass loading of 3.9 mg cm$^{-2}$. Such outstanding performance, paired with features appropriate for large-scale industrial production, enables this strategy to be prospective in the era of silicon-based batteries.

# 1. Introduction

After decades of dominance of graphite as a negative electrode in lithium-ion batteries (LIBs), silicon-based anodes are currently on the cusp of commercial success, ushering in the era of large-scale and comprehensive development.[1] Silicon (Si), with the merits of abundant reserves, ultrahigh theoretical specific capacity (4200 mAh g$^{-1}$), and attractive voltage plateau (< 0.4 V vs Li/Li$^+$), has been considered as the most prospective anode material to shine in a variety of energy storage applications including portable electronics and electric vehicles (EVs).[2] In the early stages, substantial efforts have been made in scientific research to solve the key disadvantages of Si, such as the low electrical conductivity, redundant solid electrolyte interphase (SEI) growth, and huge volume expansion (300-400%). Among numerous strategies, three types of approaches have been widely explored and recognized, including (i) the surface coating or structural embedding of Si with various carbon materials to improve electrical conductivity and interface,[3,4] (ii) the introduction of Si nanoparticles (SiNPs, < 150 nm) to minimize mechanical stress and pulverization,[5] (iii) functional binder design strategies that implant the interlayer bonding (e.g., covalent interactions, hydrogen bonding) in polymer binders to increase the mechanical strength of the entire electrode.[6,7]

However, despite numerous studies over the past decades claiming that they have fabricated Si-based anodes with a high Si content (> 50%) to greatly perform their capacity advantage (1000-3000 mAh g$^{-1}$) while achieving long-term cycling stability (> 500 cycles),[8-11] commercial LIBs used in industry have so far still been limited to a Si ratio of 5-10%.[12] In addition, most enterprises are using silicon oxides (SiO$_x$) as the anode materials, compromising the advantage of high capacity to mitigate the high electrode volume expansion problem. For example, the batteries assembled in Tesla Model 3 EVs employ less than 10% SiO$_x$ and graphite as the composite anode.[11] Samsung SDI's first-generation Si anodes, launched in 2018, had an Si content of just 2%, and the target for 2024 is only 10%.[13,14] The Si-based anode materials of NEO Battery Materials, which plans to deploy large-scale EVs after 2022, is also doped with just 5% Si.[15]

The reasons for the huge gap between scientific research and practical application can be attributed as follows. On one hand, high tap density electrode materials (or high areal active mass loading electrodes) are of great significance and necessity for commercial LIBs.[10] Compared to a loose electrode, a dense electrode possesses higher mass loading at the same electrode thickness, ensuring sufficient energy storage and outstanding areal capacity.[16] When considering device-level storage, it is feasible to minimize the relative weight of the passive components, resulting in a significant increase in gravimetric/volumetric energy and power densities. Moreover, dense electrodes can avoid unnecessary pores or voids compared to the loose electrode, thus eliminating some unwanted collateral issues such as electrolyte consumption, redundant side reactions, and massive SEI formation. Unfortunately, most research efforts remain in the laboratory stage, and the importance of this aspect is often overlooked. Until now, the majority of reported nanostructured Si-based anode materials have the tap density ranging from 0.1 to 0.5 g cm$^{-3}$, corresponding to a mass loading of less than 1.5 mg cm$^{-2}$.[10,11] Although they have reported good cycling stability under high Si ratio that effectively addresses the drawbacks of Si anodes, the potential use of such electrodes for LIB applications remains inconclusive due to the low mass loading.[17]

As a result, some researchers are beginning to restructure their study focus, with the goal of increasing the tap density of Si-based anode materials in recent years.[18-20] For instance, Tian et al. combined vacuum evaporation technique and chemical vapor deposition to fabricate magnesium-doped SiO$_x$ (SiMg$_y$O$_x$@C) microparticles with a high tap density of 1.15 g cm$^{-3}$.[18] Through a series of oxidation, acid etching, and calcination processes, Li et al. modified micro-sized Si utilizing boron doping, carbon nanotube wedging, and graphite as a robust framework, elevating the mass loading of electrodes to near commercial levels (11.2 mg cm$^{-2}$).[21] However, despite the fact that the tap density and mass loading of Si-based anode materials have been increased to some extent, considerable efforts and comprehensive consideration are still required to avoid the complexity of the preparation process and electrode structural design. This is because high-yield, economical, and scalable electrode materials synthesis processes are also considerable for the commercial development of Si-based anodes, especially for achieving the optimal balance between industrial compatibility (e.g., tap density, yield efficiency, and scalability) and electrochemical performance (e.g., mass loading, areal capacity, and cycle life).[22]

Hence, from a practical and feasible point of view, we herein present a new synthesis strategy to fabricate Si-based anode materials (blending with 20% SiNPs) with excellent electrochemical performance and industrial compatibility. Specifically, an enhanced spark plasma sintering (SPS) technique is proposed to prepare SiNPs and expanded graphite (EG) monolithic composite blocks with carbon interlayer welding (denoted as W@SG). Notably, while enhanced SPS imparts high tap density and exceptional yield efficiency advantages to W@SG anode materials, nano-welding regions assembled with the interlayer bonding structure can greatly regulate and accommodate Si expansion via mitigating the inner mechanical strain, which maintains the stable and continuous SEI layer of the W@SG anode during cycling to ensure high electrode stability. In particular, under the direct influence of both current and pressure, SPS is capable of inducing the directional distribution and arrangement of EG layers as well as the formation of closed-edge structures, thus fully encapsulating SiNPs. As a consequence, a dense packing of W@SG composite blocks can be obtained (1.68 g cm$^{-3}$), resulting in a high tap density of secondary clusters (1.12 g cm$^{-3}$). While keeping the benefits of nanostructured Si, this can also overcome its inherent limitations, such as reducing interparticle space and avoiding direct contact with the electrolyte, resulting in compact physical contact between Si and EG, a stable interface between the electrode and the electrolyte, as well as a high electrical and ionic conductivity for the W@SG anode. In addition, the introduction of interlayer welding induced by enhanced SPS endows W@SG anodes with robust mechanical protection, showing higher Vickers hardness (658 MPa) and Young's modulus (11.6 GPa) compared to the effect of binder implantation (e.g., Vickers hardness: 4-150 MPa, Young's modulus: 0.2-6 GPa). As a result, the as-prepared W@SG electrode delivers an areal capacity of 2.9 mAh cm$^{-2}$ (corresponding to 736 mAh g$^{-1}$) under a high active mass loading of 3.9 mg cm$^{-2}$ at 0.1 A g$^{-1}$ and a capacity retention of 93% after 100 cycles. Furthermore, such an electrode material synthesis approach has high scalability and yield, delivering 87 g for each 30 min process cycle. Therefore, this work provides a practical strategy for preparing industry-level Si-based anode materials, with performance and industrial compatibility benefits that are predicted to speed up commercial feasibility.

## 2. Results and Discussion

SPS technology can achieve fast and homogenous densification through Joule heating, which is applied in this work, as an appropriate electrode material processing technique for large-scale industrial production.[23] In addition, unlike conventional SPS systems, where the current is shunted greatly through the graphite die and conductive inlay, making thermal radiation the main energy source for sintering,[24,25] we employed an enhanced SPS system with a reduced resistivity of the mold and an insulating inlay foil, in which the current can only flow through the powder and thus, is the only energy source to ensure the maximized SPS effect (all the current and pressure are applied to the powder at the same time, with the whole Joule heating and densification effects) (Figure S1).[26-28] As shown in **Figure 1**a, when homogeneous SiNPs-EG composite powder (denoted as SG) is treated with an enhanced SPS process, ultra-high current flows continuously along conductive EG and the contact area between the layers, accompanied by strong pressure driving SiNPs with EG, and EG layer with EG layer to bond tightly. Subsequently, the EG layers would be aligned perpendicular to the applied pressure as a result of uniaxial pressing, with robust and durable interlayer welding and closed-edge structures that seamlessly encapsulates and locks SiNPs inside, which contributes to stabilizing the Si anode throughout the lithiation and delithiation processes. Notably, the as-prepared W@SG differs from most reported Si/carbon coating structures,[29-31] which are merely physical surface covering or blending, with very weak interactions between Si and carbon materials.

In order to observe the changes in morphology and structural characteristics of the as-prepared active materials before and after enhanced SPS treatment, scanning electron microscopy (SEM), X-ray photoelectron spectroscopy (XPS), and nitrogen adsorption-desorption characterizations are illustrated in Figure 1b-g. After ball milling, the as-obtained SG exhibits a homogeneous black powder, in which well-dispersed SiNPs are embedded between EG layers (Figure 1b and Figure S2c, d). After enhanced SPS treatment, a monolithic W@SG block is synthesized with a vast majority of SiNPs anchored and encapsulated inside, that can be further milled into micron-sized secondary clusters (Figure 1c and Figure S2e, f). According to XPS spectra (Figure 1d) and the atomic percentage analysis (Figure S3a), there is a substantial drop in Si 2p peak strength and Si atomic ratio, indicating that the SG and W@SG surfaces are covered and occupied by EG layers (as the penetration depth of X-ray photoelectrons is a few nanometers).[31,32] This carbon coverage structure can not only effectively improve the electrical conductivity of the electrode material, but also greatly avoid direct contact between SiNPs and electrolyte, reducing unnecessary side reactions and repeated unstable SEI formation.[10] In addition, the cross-section SEM images of W@SG shown in Figures 1e and f display the directionally aligned EG layers, and the edges between the layers are all welded together, making SiNPs perfectly wrapped inside. When SPS pressure is reduced further, similar structures can be maintained in W@SG-S80 (SPS, 800 °C, 80 MPa) and W@SG-S40 (SPS, 800 °C, 40 MPa). As illustrated in Figure S4, the SiNPs and EG layers are tightly adhered to each other in the SPS-treated samples, preserving most of the closed-edge structure, whereas the hot-pressed sample (SG-H40, HP, 800 °C, 40 MPa) has an open and disordered structure. The SiNPs in SG-H40, in particular, can be seen plainly in the crevices between the EG layers. Moreover, according to the results of nitrogen adsorption/desorption isotherms shown in Figure 1g and Table S1, the total pore volume and average pore size of W@SG are considerably decreased compared to that of SG, resulting in a lower specific surface area of 37.8 $m^2$ $g^{-1}$. As a result of the enhanced SPS effect, the formation of dense and ordered coating structures with unique closed edges is demonstrated, which is beneficial for providing uniform electric field distribution and stabilizing the interface between the W@SG anode and electrolyte.

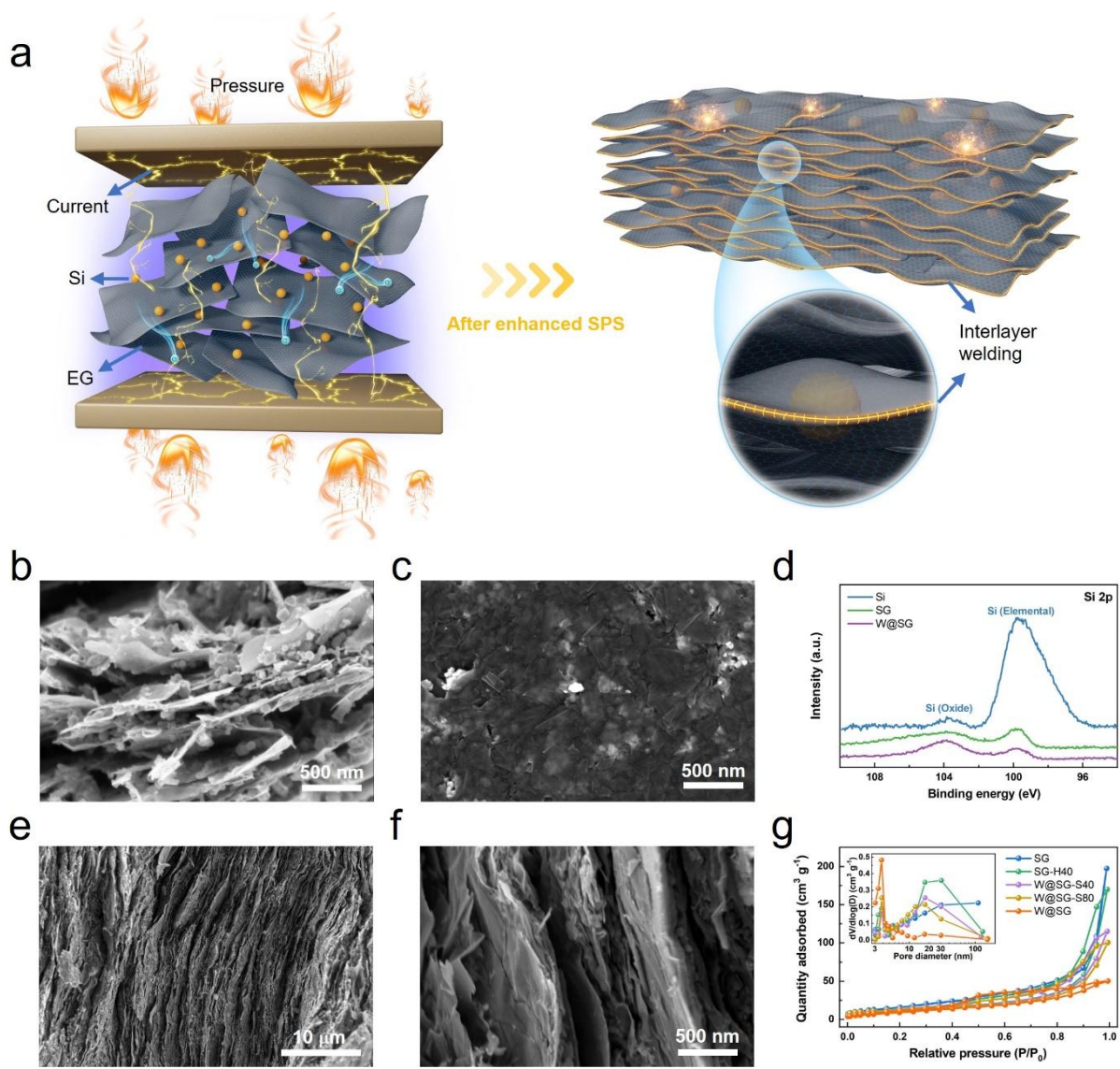

**Figure 1.** (a) Schematic diagram of W@SG obtained from SG via enhanced SPS technology. (b, c) Planar SEM images of (b) SG and (c) W@SG. (d) XPS spectra comparison of Si 2p of bare Si, SG, and W@SG. (e, f) Cross-sectional SEM images of W@SG with low and high magnification. (g) Nitrogen adsorption/desorption isotherms of the as-prepared materials, inserted with Barrett-Joyner-Halenda (BJH) pore size distribution curves.

Furthermore, the detailed microstructure and physical properties, including transmission electron microscopy (TEM), energy dispersive spectrometer (EDS) mapping, indentation tests, conductivity and tap density characterizations of the as-prepared anode materials, are analyzed in **Figure 2**. As shown in Figure 2a and Figure S5, TEM and EDS results of SG further confirm that SiNPs are well embedded and wrapped in peeled EG layers. More importantly, the EG layers in SG show a flat morphology with clear and smooth edges, while in W@SG they present closed-edge or rolled-up structures (with more dark strip morphology), which we hypothesized could be ascribed to the interlayer welding effect caused by enhanced SPS (Figures 2b and c, and Figure S6). Notably, the welded EG is similar to structures induced by radiation (e.g., neutron, electron, and ion) with the formation of interlayer bonding on the c-axis reported in the literature, implying that defects and reconstruction occur in the material under continuous pressure and high current, especially at the edges and intersections of EG layers.[33-35] These interlayer bonds, consisting of a small number of interstitial atoms excited by a continuous strong current, can produce a pinning effect on mobile dislocations within the EG basal planes, thereby reducing elastic and plastic deformation under an applied stress.[36]

To investigate how this structure affects the mechanical properties and find indirect evidence of the atomic structure, we performed the indentation test as shown in Figures 2d and e. Compared to other samples (W@SG-S80, W@SG-S40, and SG-H40), W@SG exhibits the strongest mechanical stability with a Vickers hardness of 658 MPa and a Young's modulus of 11.6 GPa, resulting from the formation of interlayer welding within the EG structure. With the raising of SPS pressure, both Vickers hardness and Young's modulus of the samples show an increasing trend, indicating that it would generate more interlayer bonds to enhance the nano-welding effect under increasingly harsh SPS conditions. Moreover, these values are much higher than those reported in the Si-based anode literature (Figure 2g and Table S2).[37] [7] [38] [39] [40] [41] [42] [43] It's worth noting that most of the approaches for improving mechanical properties in Si-based anodes involve the introduction of hard or soft binders to generate hydrogen bonding, covalent bonding, or strong intermolecular interactions, whereas our system allows directly processing the electrode material to make itself more mechanically stable with a robust closed carbon shell protection. In addition, this sealed carbon interlayer-welding structure provides W@SG an extremely high conductivity of 320 S cm$^{-1}$ (compared to reported Si-based composites, e.g., 1.2 S cm$^{-1}$ for carbon/Si,[44] 67 S cm$^{-1}$ for graphene/Si,[45] 211 S cm$^{-1}$ for carbon nanotubes/Si[32]), as well as a high tap density (monolithic block: 1.68 g cm$^{-3}$, secondary clusters: 1.12 g cm$^{-3}$) (Figure 2f), and better thermal stability than that of SG (Figure S8). More importantly, this confirms that the enhanced SPS technique can considerably improve the tap density of the anode material to a comparable commercial level based on nano-sized Si as opposed to the reported methods directly using micro-sized Si (Figure 2h and Table S3).[9] [46] [47] [19] [29] [48] [49] [8] [20] [50] [51] [18] As a consequence, there are three main benefits: (i) The W@SG anode can fully exploit the advantages of nano-sized Si, such as the minimum crushing level after long-term cycling. (ii) With the protection of the closed and dense carbon shell, the W@SG anode can significantly overcome the deficiencies of nano-sized Si, such as avoiding the formation of excessive SEI caused by direct contact between SiNPs and electrolyte, and improving the conductivity through uniform and continuous electron transportation channels provided by the welded EG layers. (iii) The high tap density can give the W@SG anode a high electrode loading comparable to an industrial level, which serves to boost the energy density of the battery and make it commercially viable.

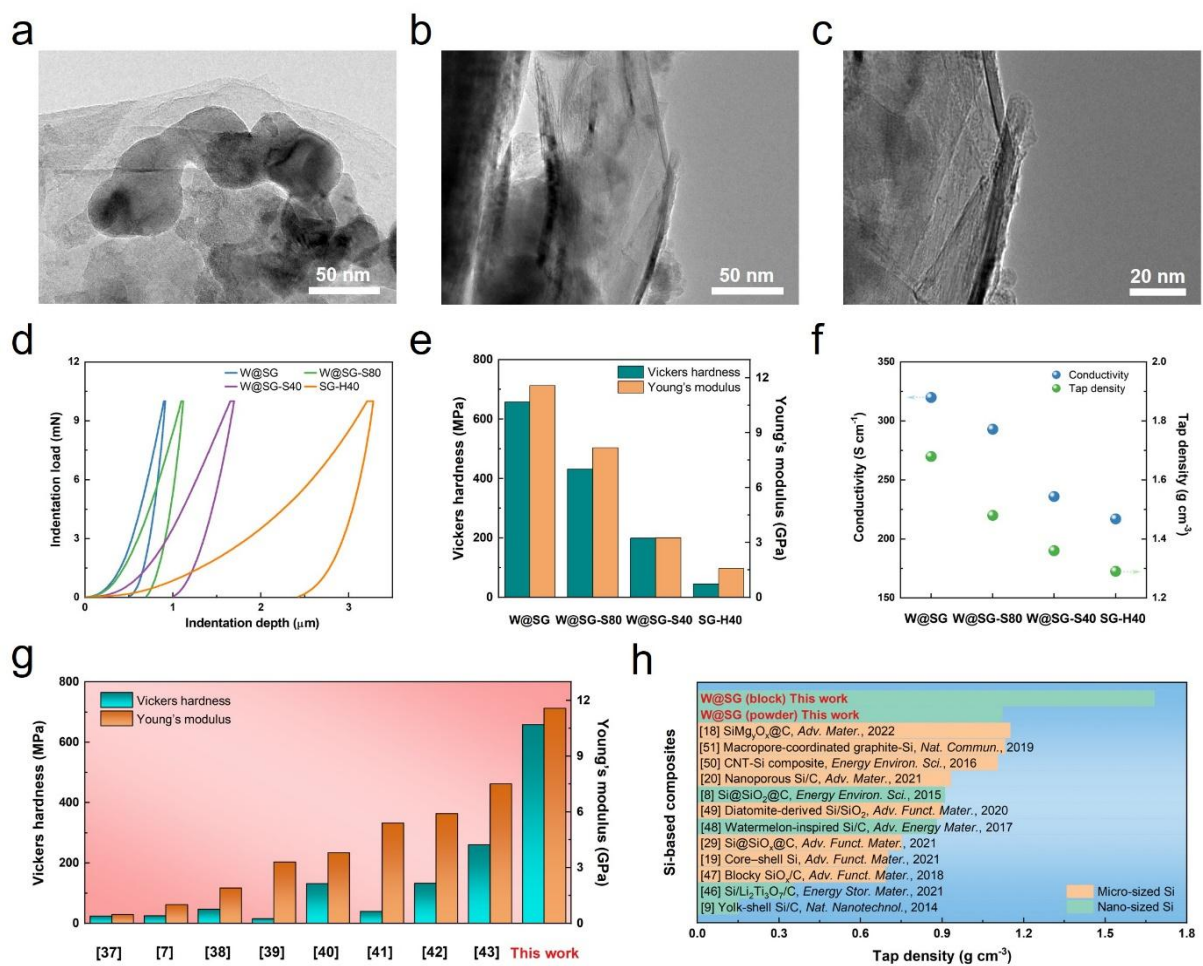

**Figure 2.** The microstructure and physical properties of the as-prepared anode materials. (a) TEM image of SG. (b, c) TEM images of W@SG. (d) Indentation tests on various Si-based anode materials, including W@SG, W@SG-S80, W@SG-S40, and SG-H40. (e) Vickers hardness and Young's modulus results of the as-prepared anode materials from the indentation test. (f) Tap density and conductivity results of the as-prepared anode materials. (g) Comparison of W@SG with reported Si-based electrodes in Vickers hardness and Young's modulus. (h) Comparison of W@SG with reported Si-based electrodes in tap density.

In addition, we analyze the atomic structure changes of the material after enhanced SPS treatment (**Figure 3**). Similar to the irradiated graphite,[34] under an ultra-harsh environment of 800 °C and 120 MPa, SG will be continuously subjected to a current of up to 4 kA and a power of 20 kW, accompanied by the reconstruction of its atomic structure (Figure 3a).[35,52] Firstly, a certain of functional groups (Figure S3b) and impurities are removed under high current irradiation, and then a large number of structural defects (e.g., vacancy), dislocations (e.g., a change in the arrangement of atoms) and dangling bonds are progressively produced in the wrapped carbon layer of SG.[53] Subsequently, the dangling bonds will be re-bonded with each other or with the defective atoms, forming the pinning effect.[34] Finally, W@SG with a substantial number of structural defects and interlayer bonding is formed, which corresponds to the XPS spectra with an increased amount of $sp^3$ hybridized carbon atoms (Figures S3c and d). Moreover, these structural changes can also be confirmed by the results of Raman spectra and X-ray diffractometer (XRD) characterization. As shown in Figures 3b and c, the intensity ratio between D-peak and G-peak ($I_D/I_G$) increases from 0.16 for pristine EG to 0.62 for W@SG, indicating the surge of disorder and structural defects. Besides, the $I_D/I_G$ ratio of W@SG-S40 is larger than that of SG-H40, revealing that SPS effect is more conducive to the formation of such disordered and interlayer bonding structures than thermal effect. As the pressure increases, the SPS effect becomes more pronounced, resulting in more structural defects and disorder. Besides, XRD characterization is performed in Figures 3d and e to detect the crystalline phase of all samples. Obviously, the as-prepared samples all have high Si crystallinity, with five characteristic diffraction peaks at 28.5°, 47.3°, 56.2°, 69.1° and 76.4°, corresponding to the (111), (220), (311), (400) and (331) planes of crystalline Si (JCPDS No. 27-1402), while two peaks near 26.5° and 54.6° are in agreement with the (002) and (004) planes of graphite (JCPDS No. 41-1487), respectively (Figure 3d). Interestingly, it can be observed from Figure 3e that the position of the (002) peak shifts slightly to lower diffraction angles, indicating an increased interlayer distance trend with the enhancement of the SPS effect. The striking point is that the tap density of the material increases with densification, while the interlayer distance does not decline but rather climbs. This may be attributed to the generation of the interstitial defects produced by enhanced SPS, leading to the expansion of graphite crystallites along the c-axis direction, which also confirms the existence of interlayer bonds.[36,54] In a word, this atomic structure with the interlayer bonding can limit the slipping motion of EG layers and considerably enhance the interlayer interaction, which is beneficial to improving the overall mechanical stability of the W@SG anode (e.g., hardness, tensile strength, elastic modulus). In addition, increasing layer spacing makes it easier to insert/extract ions (e.g., $Na^+$, $K^+$, $Li^+$), with the opportunity to dramatically promotes reaction kinetics.[55,56]

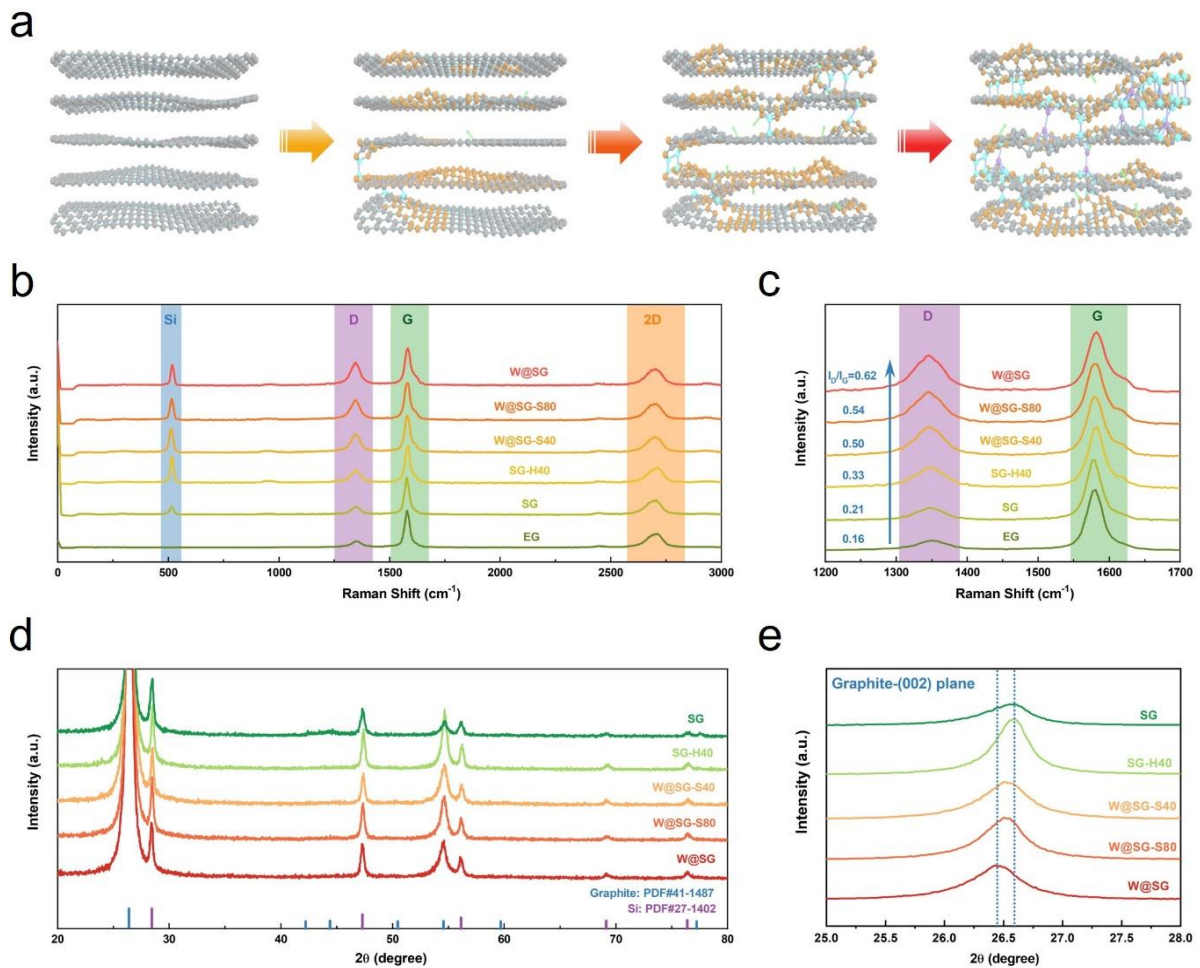

**Figure 3.** Analysis of atomic structure. (a) Demonstration of atomic structure changes of EG treated by enhanced SPS technology. (b, c) Raman spectra of the as-obtained anode materials. (d, e) XRD patterns of the as-obtained anode materials.

To further investigate how this unique structure affects the electrochemical performance of LIBs, a series of electrochemical characterizations are performed on the as-prepared Si-based anodes, as shown in **Figure 4**. The cyclic voltammetry (CV) curves are used to explore the kinetics mechanism of the W@SG anode during the charging/discharging process (Figure 4a). In the first cathodic scan, two broad reduction peaks are located at approximately 0.7 and 0.9 V, indicating the formation of an SEI layer on the anode, and these two peaks will vanish in subsequent cycles.[56,57] After that, two additional cathodic peaks appear at 0.16 and 0.20 V, respectively, due to the formation of $Li_xC$ compound and $Li_xSi$ alloy, which also trigger the peak below 0.1 V. In addition, there are two main peaks located at 0.29 and 0.51 V in the first anodic scan, which reflect the delithiation process. Subsequently, they would progressively evolve into four peaks, with two peaks at 0.23 and 0.28 V depicting the transformation from the $Li_xC$ phase to C, and two peaks at 0.32 and 0.51 V representing the conversion from the $Li_xSi$ phase to Si.[56,57] Obviously, both Si and EG are involved in the electrochemical response of the

W@SG electrode, revealing that the overall lithium storage is a synergistic effect of them. In addition, as shown in Figure S9a, the peak strength of Si in the SG anode decreases after the 6th cycle, while that of W@SG anode still maintains an upward trend, indicating that some crystalline Si still exists in the anode, requiring a certain cycle to make full use of its capacity, which reflects the protective effect of the interlayer-welding structure on Si. Moreover, similar trends have been observed from the CV curves of SPS-treated samples (W@SG-S80 and W@SG-S40), indicating that the good synergistic effect of Si and C jointly promotes electrochemical reactions (Figure S9).

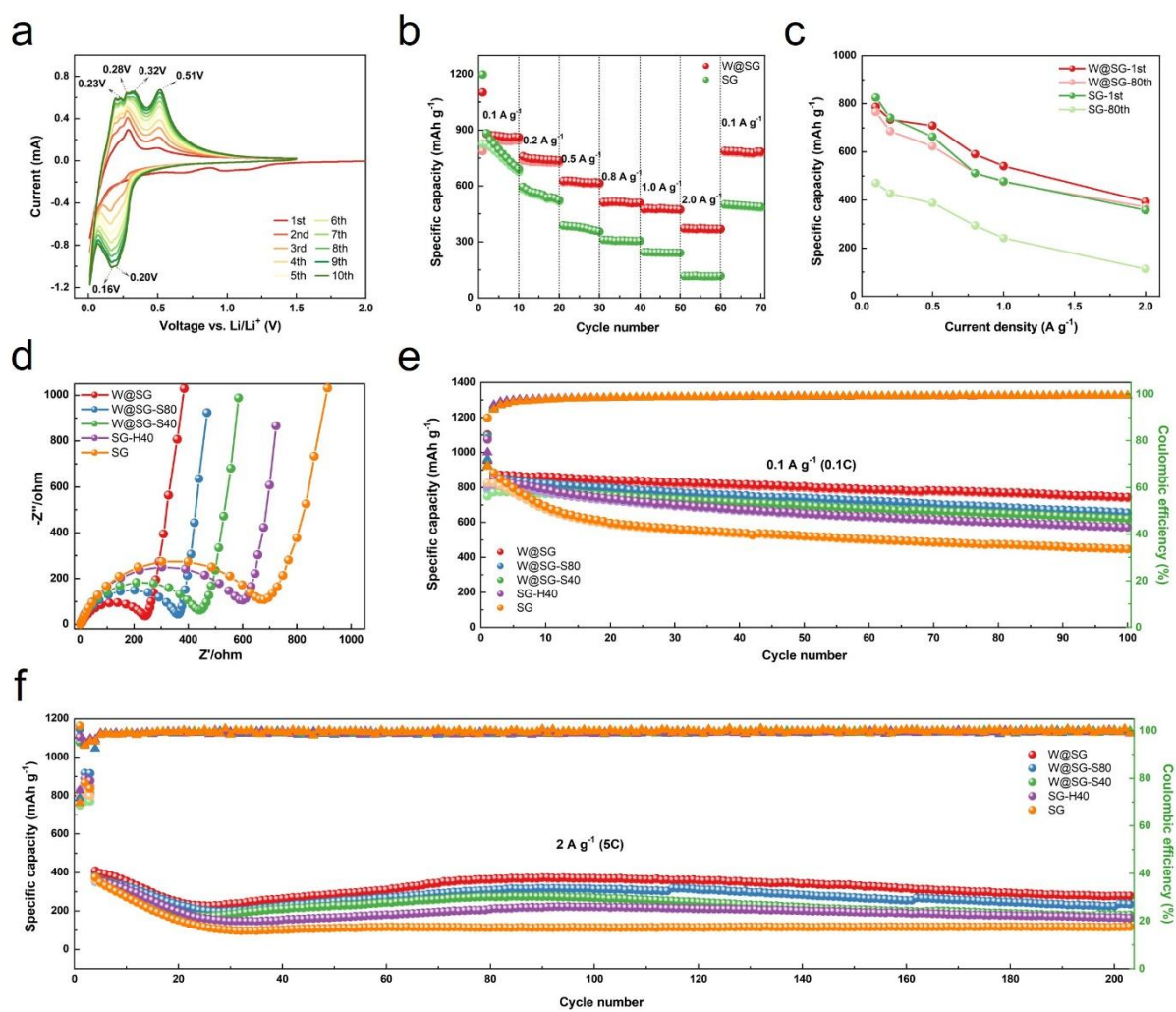

**Figure 4.** Electrochemical performance of the as-obtained anodes. (a) CV curves of the W@SG anode at a scanning rate of 0.1 mV s$^{-1}$. (b) Rate performance of the W@SG and SG anodes at various current densities. (c) The specific capacities comparison of the W@SG and SG anodes after the 1st and 80th cycles at different current densities. (d) EIS curves of the fresh W@SG, W@SG-S80, W@SG-S40, SG-H40, and SG electrodes. (e) The cycling stability of different electrodes at 0.1 A g$^{-1}$. (f) Long cycling performance of different electrodes at 2 A g$^{-1}$.

As an important reflection form of the Li$^+$ intercalation/deintercalation kinetics, the rate capability is greatly affected by the structural properties of electrode materials, such as the electrical and ionic conductivities. Therefore, the rate performance of W@SG and SG electrodes is further investigated with current densities varied from 0.1 to 2 A g$^{-1}$ in Figure 4b. Although W@SG and SG electrodes possess comparable capacities at 0.1 A g$^{-1}$ for the first few cycles, as the current density is increased, the capacity and stability variations become much more visible. In particular, the specific capacity of the SG electrode plummets linearly at low current densities, and while it becomes somewhat stable at high current densities, its capacity values are quite low. By contrast, the W@SG anode can deliver higher and more stable reversible capacities of 844, 730, 621, 510, 475, and 371 mAh g$^{-1}$ for 10 consecutive cycles at current densities of 0.1, 0.2, 0.5, 0.8, 1, and 2 A g$^{-1}$, respectively. When the current density is switched back to 0.1 A g$^{-1}$, the specific capacity of the W@SG anode recovers to 781 mAh g$^{-1}$, superior to that of the SG anode (497 mAh g$^{-1}$), demonstrating the outstanding rate capability. Furthermore, the specific charging capacities of W@SG and SG anodes at the 1st and 80th cycles are compared in Figure 4c, to investigate the cycling stability of electrodes at all current densities. Obviously, the reduction range of the specific capacities in the W@SG anode is significantly lower than that of SG at various rates after 80 cycles. In particular, the W@SG anode can achieve an ultra-high capacity retention of 97.4% after 80 cycles at a low current density of 0.1 A g$^{-1}$, as well as an incredible retention of 94.2% at a raised current density of 2 A g$^{-1}$, which further confirms the superior rate and cycling performance of the W@SG electrode (Figure 4c and Figure S12, Table S4).

Subsequently, the electrochemical impedance spectroscopy (EIS) measurements of the W@SG, W@SG-S80, W@SG-S40, SG-H40, and SG electrodes are performed in Figure 4d, with relevant equivalent circuit model and fitting data shown in Figure S13 and Table S5. Compared to other electrodes, the fresh W@SG anode has the lowest charge transfer resistance ($R_{ct}$) of 216 Ω, indicating the best conductivity, which can be attributed to the uniform and stable electric field environment of W@SG provided by the tightly closed interlayer-welding EG shell (Figure 4d). After the 1st cycle, the W@SG electrode develops a stable and homogenous SEI layer with a low resistance of 2.1 Ω ($R_{SEI}$), showing that there is less contact area between SiNPs and electrolyte and fewer side reactions (Figure S13 and Table S5). Moreover, the $R_{ct}$ of the W@SG electrode is significantly reduced to 6.1 Ω, much smaller than that of SG (132.5 Ω), because of an increased electrochemical conductivity and an effective dynamic environment formed under uniform benign SEI coverage. After 100 cycles, the W@SG anode still maintains a similar $R_{SEI}$ of 4.1 Ω, indicating that SiNPs did not undergo severe crushing, and the SEI layer retained stable morphology and function to suppress the decomposition of electrolyte. Meanwhile, the $R_{ct}$ is further lowered to 3.5 Ω, suggesting that the surface kinetics has been improved after multiple cycles as a result of enhanced wetting status and the resilient and consistent interlayer-welding EG shell. As a consequence, the robust and directional interlayer-welding EG shell provides excellent electronic and ionic conductivities for the W@SG anode to generate faster dynamic reactions, thereby enabling the enhanced rate capability, as well as fully and consistently exploiting its capacity advantage.

The cycling stability of the as-prepared Si-based anodes at 0.1 A $g^{-1}$ and 2 A $g^{-1}$ is investigated in Figure 4e and f. The W@SG electrode consistently shows the highest reversible capacity at 0.1 A $g^{-1}$ (corresponding to 0.1C) throughout the stage, whose stable capacity of 739 mAh $g^{-1}$ is recorded after 100 cycles, with a high retention of 94% corresponding to a low capacity decay rate of 0.06% per cycle, and the Coulomb efficiency (CE) startlingly climbs to 99.3% (Figure 4e and Table S6). In sharp contrast, the SG electrode only has a specific capacity of 446 mAh $g^{-1}$ after 100 cycles at 0.1 A $g^{-1}$, showing fast capacity fade with only 53.9% capacity retention. In addition, when the charging/discharging time is further extended (Figure S14), an ultra-high CE of 99.8% in the W@SG electrode can be achieved, which outputs a stable capacity of 495 mAh $g^{-1}$ after 300 cycles (with a small decay rate of 0.109% per cycle), significantly higher than that of W@SG-S80 (459 mAh $g^{-1}$), W@SG-S40 (410 mAh $g^{-1}$), SG-H40 (383 mAh $g^{-1}$), and SG electrodes (333 mAh $g^{-1}$). More importantly, at the low current densities of 0.1 A $g^{-1}$ and 0.2 A $g^{-1}$, the SPS-treated samples exhibit a trend of steady reduction in capacity, while the SG-H40 and SG electrodes both decrease significantly in the initial stage, with large capacity attenuation, which directly indicates that the robust carbon interlayer-welding structure induced by enhanced SPS is beneficial to stabilizing the Si anode, thus improving the cycling performance. When the current density is further increased to 2 A $g^{-1}$ (corresponding to a fast charge/discharge rate of 5C), as shown in Figure 4f, the W@SG electrode can retain 93% capacity after 100 cycles, and 70.4% capacity even after 200 cycles, which is much higher than other electrodes (W@SG-S80: 66.9%, W@SG-S40: 49.6%, SG-H40: 45%, SG: 31.6%). Besides, the variation trend of capacity decreasing and then rising in the first 100 cycles can be found, especially for the W@SG electrode, which can be attributed to the fact that the secondary blocks have a large particle size, requiring a certain number of cycles to fully exploit the capacity of the embedded SiNPs.[57] This further supports the notion that SiNPs are adequately protected, as both the high-C rate cycling and the small sweep speed CV cycling require a specific number of cycles to completely perform the capacity performance of Si.

To verify the origin of the enhanced cycling stability of W@SG, we investigated the

morphology and structural changes of W@SG and SG electrodes before and after cycling, as shown in **Figure 5**. According to the cross-section SEM images (Figure 5a), the thickness of the W@SG electrode hardly fluctuates after 100 cycles, with a change value of 5.1 μm, corresponding to a low electrode swelling of 14.3%, which is close to the commercial requirement (< 10%).[11] In the case of the SG electrode, however, the thickness change is recorded as 26.8 μm, with a catastrophic electrode swelling of 82.7%. Besides, it is evident that the W@SG electrode maintains a flat, continuous, and dense morphology after cycling, and the SiNPs are well wrapped by EG sheets with strong physical and electrical contact, allowing them to preserve good integrity without pulverization (Figures 5c-e). In sharp contrast, the morphology and structure of the SG electrode are plainly disrupted during the lithiation/delithiation processes, resulting in many cracks and cavities on the electrode surface, as well as severely deformed and fractured SiNPs (Figures 5f-h). These further indicate that the well-designed interlayer-welding structure provides robust mechanical protection to regulate and accommodate Si expansion, thereby mitigating the inner mechanical strain and sustaining the stable morphology/structure of the W@SG electrode during cycling, which creates the prerequisite for maintaining stable electrical contact and electrochemical interface, resulting in weak electrode swelling and excellent cycling stability. In more detail, because the SiNPs are well-encapsulated inside the welded EG layer, the SEI layer of the W@SG electrode is specifically produced outside each of the secondary blocks. During the lithiation process, in order to adapt to the expansion of SiNPs, the robust mechanical strength of W@SG blocks (ultra-high Young's modulus and Vickers hardness) caused by the pinning effect coming from the interlayer bonding makes the anode only produce very low strain under extreme stress. Following delithiation, W@SG blocks undergo entropic repulsion to revert to their initial high-energy state. Therefore, it won't impair the overall morphology/structure of the whole block, thus ensuring a continuous and stable SEI.[58] In contrast, the SiNPs in the SG electrode form separate SEI layers that rupture after multiple cycles.

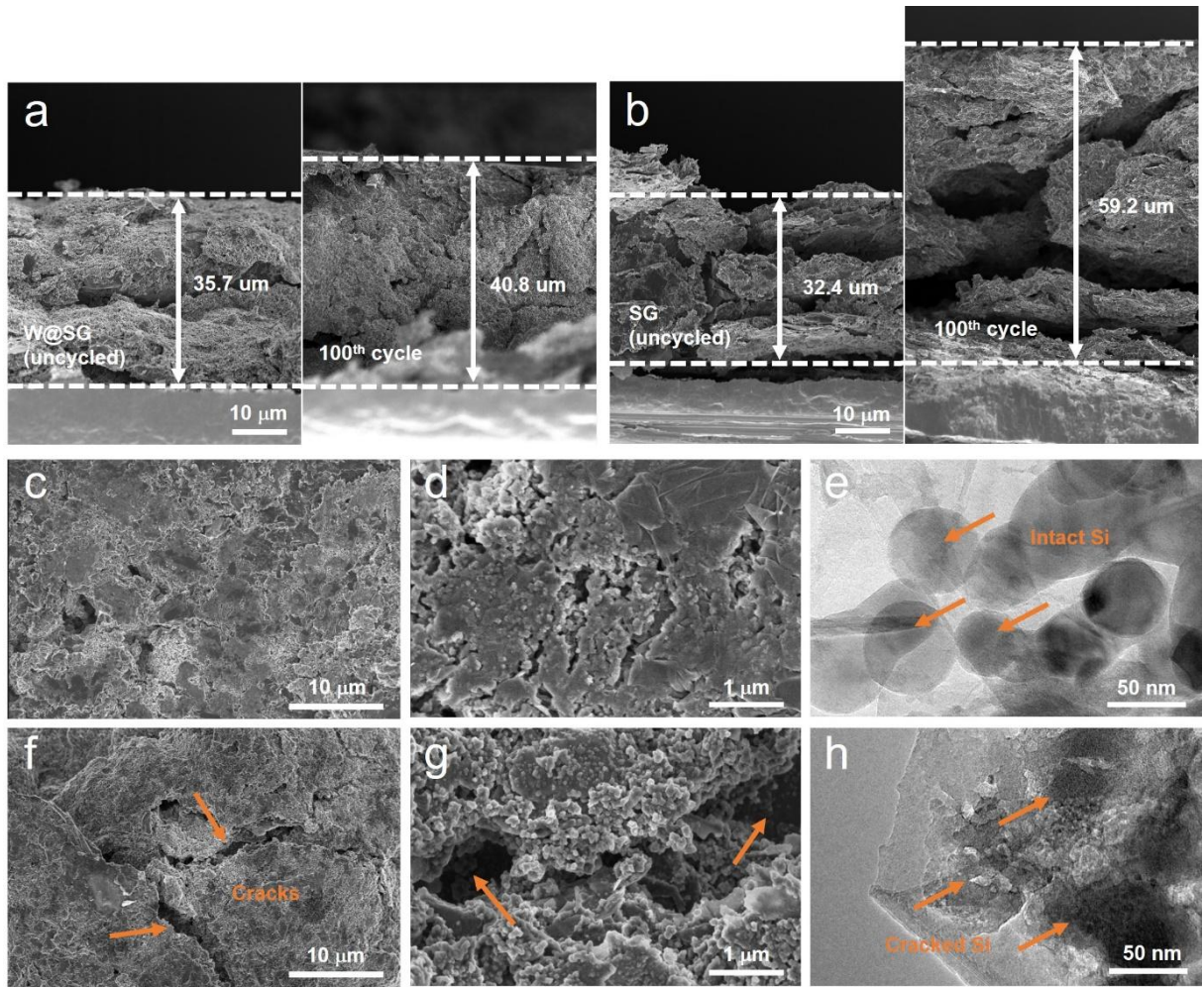

**Figure 5.** (a) Cross-section SEM images of the W@SG electrode before and after 100 cycles. (b) Cross-section SEM images of the SG electrode before and after 100 cycles. (c-e) Planar SEM and TEM images of the W@SG electrode after 100 cycles. (f-h) Planar SEM and TEM images of the SG electrode after 100 cycles.

To demonstrate the potential for practical applications, the electrochemical performance of the W@SG anode under commercial loading level and the industrial compatibility of the preparation process are strictly evaluated. As shown in **Figure 6**, with a high electrode mass loading of 4.8 mg cm$^{-2}$ (corresponding to an active mass loading of 3.9 mg cm$^{-2}$), the W@SG anode maintains outstanding capacity and cycling performance. In particular, the cycling stability of the W@SG electrode at a current density of 0.1 A g$^{-1}$ is evaluated as shown in Figure 6a. Similar to the W@SG-1.7 mg cm$^{-2}$ electrode, a steady curve of specific capacity with the cycle number can also be detected for the W@SG-3.9 mg cm$^{-2}$ electrode at 0.1 A g$^{-1}$, showing a stable specific capacity of 632 mAh g$^{-1}$ after 100 cycles, with a high-capacity retention of 93% and an average CE of 98.5%. Meanwhile, the W@SG-3.9 mg cm$^{-2}$ electrode has the ability to operate continuously for 150 cycles at 0.2 A g$^{-1}$ while still maintaining a competitive capacity level (491 mAh g$^{-1}$), demonstrating good stability (Figure S15). In addition, the W@SG-3.9 mg

cm$^{-2}$ electrode can deliver stable specific charging capacities of 736, 636, 503, 371, and 182 mAh g$^{-1}$ at current densities of 0.1, 0.2, 0.5, 1, and 2 A g$^{-1}$, respectively (Figure 6b). Notably, when the mass loading is increased by 2.5 times, the W@SG electrode can still produce a high specific mass capacity (736 mAh g$^{-1}$) at 0.1 A g$^{-1}$, with 87.2% retention compared to that at low areal loading (844 mAh g$^{-1}$), indicating that the ion and electron transport is not significantly affected, and an unimpeded and vigorous internal dynamic can be maintained. At the same time, the areal capacity of the W@SG electrode has skyrocketed to 2.9 mAh cm$^{-2}$, with an increase of 207%. Moreover, it should be noted that the output capacity of the W@SG electrode (182 mAh g$^{-1}$) is still about 5 times higher than that of commercial graphite (~36 mAh g$^{-1}$) at a high current density of 2 A g$^{-1}$.[56] More importantly, it is well known that achieving high areal capacitance and sustained cycle life at large mass loading has long been a significant indicator of industrial feasibility determination. In comparison to numerous reported Si-based anodes, the W@SG electrode, benefiting from unprecedented advantages in well-designed interlayer-welding structure, has absolute competitiveness in terms of loading level (3.9 mg cm$^{-2}$), areal capacity (2.9 mAh cm$^{-2}$), and cycle life (93% after 100 cycles), which is expected to shine in the coming era of Si-based batteries (Figure 6c and Table S7). [59] [60] [29] [61] [19] [62] [47] [48] [63] [64] [30] [50] [65] [66] [67] [9]

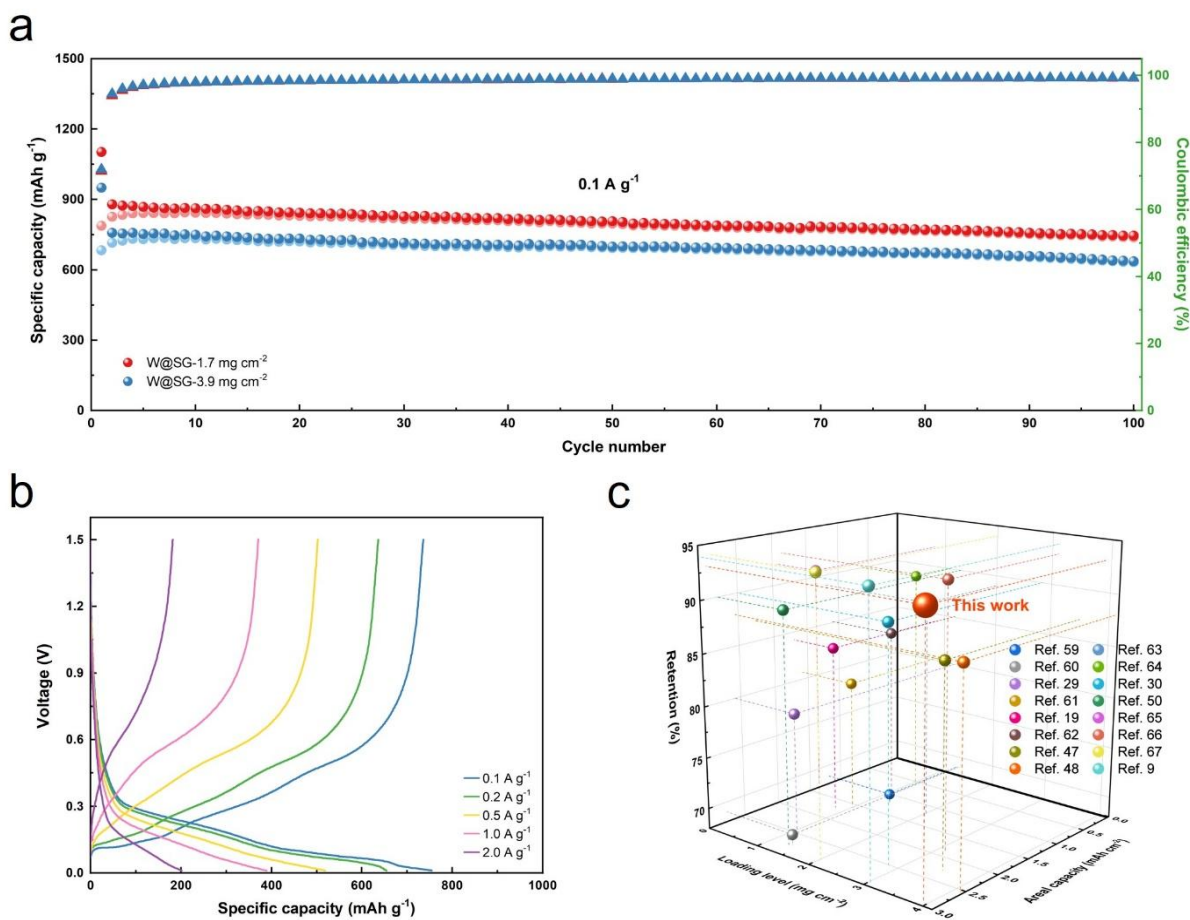

**Figure 6.** Electrochemical performance of the W@SG electrode under the active mass loading of 3.9 mg cm$^{-2}$. (a) The cycling stability of the W@SG-3.9 mg cm$^{-2}$ and W@SG-1.7 mg cm$^{-2}$ electrodes at 0.1 A g$^{-1}$. (b) Voltage profiles of the W@SG-3.9 mg cm$^{-2}$ electrode at various rates from 0.1 to 2 A g$^{-1}$ with a potential window between 0.01 and 1.5 V (vs. Li/Li$^+$). (c) Comparison of the W@SG-3.9 mg cm$^{-2}$ electrode with reported Si-based anodes in active mass loading, areal capacity, and cycle life.

Last but not least, we demonstrated the high scalability of W@SG composite monolithic blocks as shown in Figure S16. In particular, the diameter of the prepared W@SG block can be controlled by selecting different sizes of molds; the thickness can be regulated by the pressure and mass loading; and the number of W@SG blocks produced at one time can be controlled by tool cavity designs (e.g., parallel, serial, or parallel-serial alignment).[23] The largest W@SG monolithic block, for example, has a diameter of 10 cm and a thickness of 6.55 mm, providing an ultra-high output of 86 g. Therefore, as a feasible, scalable, and high-yielding electrode material preparation process, the enhanced SPS technology is capable of being used in large-scale industrial production of W@SG blocks.

## 3. Conclusion

In summary, we have developed a high-yield, scalable, and industrial-compatible synthesis process for W@SG monolithic blocks to overcome the limitations caused by large electrode loading and Si volume expansion. Enhanced SPS technology has induced the generation of directional alignment, compact coverage, and closed-edge structures by unique interlayer welding, enabling a vigorous dynamic environment with improved ionic and electronic conductivities. Besides, the nano-welding EG shells assembled with interlayer bonding provide robust mechanical protection to regulate and accommodate Si expansion, with unprecedented Vickers hardness (658 MPa) and Young's modulus (11.6 GPa), as well as industrially available tap density (1.68 g cm$^{-3}$). In particular, the issue of Si volume expansion can be effectively alleviated by releasing large amounts of stress under limited strains, preserving the integrity of the SEI layer on the W@SG electrode. As a consequence, the W@SG electrode exhibits outstanding capacity, rate, and cycling performance under the synergistic effect of EG and Si, with a stable reversible capacity of 844 mAh g$^{-1}$, as well as 94% capacity retention and 14.3% electrode swelling after 100 cycles. When the areal active mass loading is further increased to 3.9 mg cm$^{-2}$, its areal capacity (2.9 mAh cm$^{-2}$) and cycle life (93% after 100 cycles) stand out from numerous reported Si-based anodes. Furthermore, this synthesis strategy demonstrates the advantages of high yield efficiency and excellent scalability, which can output up to 86 g of electrode materials in one process. Therefore, this work provides a unique and practical strategy for mass production of Si-based anode materials with the benefits of high areal capacity and cycling stability, aiming to hasten the commercialization of large-scale Si-based batteries.

## Supporting Information

Supporting Information is available from the Wiley Online Library or from the author.

## Acknowledgements

This work was supported by the National Research Foundation, Prime Minister's Office, Singapore, under its Competitive Research Programme (CRP award number NRFCRP22-2019-8), NRF Investigatorship (NRFI award number NRF-NRFI2018-8) and Medium-Sized Centre Programme.

## Conflict of Interest

The authors declare no conflict of interest.

## Data Availability Statement

The data that support the findings of this study are available from the corresponding author upon reasonable request.

# Supporting Information

**Experimental section**

**Materials**

Expanded graphite powder (EG, 99%, 100 mesh) was purchased from Qingdao Yanxin Graphite Products Co., Ltd. Silicon nanopowder (Si, >98%, 30-50 nm, laser synthesized) was purchased from US Research Nanomaterials, Inc. Conductive carbon black (Super P), the binder materials including sodium carboxymethyl cellulose (CMC, molecular weight of 400 000) and styrene-butadiene rubber (SBR), were purchased from MTI Corporation. The electrolyte (purity: 99.9%) contained 1 M lithium hexafluorophosphate (LiPF$_6$) in a mixture of (1:1 vol%) ethylene carbonate (EC) and diethyl carbonate (DEC), with 10 wt% fluoroethylene carbonate (FEC) as additive, was purchased from Solvionic. All reagents were used without further purification.

**Preparation of SG**

In order to obtain highly uniformly dispersed composite powder, a two-step ball mill method was used in this experiment. 0.4 g of Si nanoparticles and 1.6 g of EG were dispersed in isopropanol solution (50 mL) at a mass ratio of 1:4, and then transferred to the ball mill pot after 2 h ultrasonication. Rolling ball mill (MSK-SFM-14-IVS, Hefei Kejing Materials Technology Co., Ltd) was used to perform pre-dispersion at a speed of 800 rpm for 18 h. After drying, the collected mixed powder was re-dispersed to isopropanol for high-energy ball milling (HEBM, PM200, Retsch) at 400 rpm and 12 h. Finally, homogeneous Si-EG composite (SG) was obtained after vacuum drying at 80 °C for 24 h.

**Preparation of W@SG**

The SG powder was treated in an enhanced spark plasma sintering (SPS) system (KCE®-FCT H-HP D 25-SD, FCT Systeme GmbH) to induce the formation of interlayer-welding. With its ultra-high operating current (0-10 kA) and precise force control, the enhanced SPS system can output 0-60 kW of power and 20-120 MPa of pressure for rapid, homogeneous sintering. Here, 1.5 g of SG powder was placed in a cobalt-doped tungsten carbide (WC) mold, with an insulating mica foil on the side and conductive graphite foils on the upper/lower sides to ensure that the plasma current only applies to the powder during SPS operation. After assembled in SPS chamber, an applied temperature of 800 °C, a pressure of 120 MPa and a holding time of 30 min were carried out under vacuum condition. After cooling, SPS-processed SG block with interlayer-welding structure (W@SG) was prepared, whose Si content was 20 wt%. The monolithic W@SG block was then crushed into fine particles by rolling ball mill machine for subsequent electrochemical testing and characterization. For comparison, the other samples prepared under different SPS pressures were denoted as W@SG-S80 for a pressure of 80 MPa, W@SG-S40 for a pressure of 40 MPa. To highlight the benefits and effect of SPS, the sample prepared at a pressure of 40 MPa by traditional hot press (HP) was defined as SG-H40 (other preparation parameters remained unchanged).

**Materials characterization**

The morphologies of the samples were observed by field emission scanning electron microscopy (SEM, VERIOS 460, FEI). Further detailed characterization was conducted by high-resolution transmission electron microscopy (TEM, JEM-3010, JEOL) equipped with energy dispersive spectrometer (EDS). The chemical composition and structural characteristics were analyzed by an X-ray diffractometer (XRD, D8 Advance, Bruker) equipped with a Cu K$\alpha$ radiation, Raman microscope (Alpha 300R, WITec) with a 532 nm laser, X-ray photoelectron spectroscopy (XPS, Escalab 250Xi, Thermo Fisher Scientific) equipped with a monochromatic Mg K$\alpha$ source. After outgassed at 300 °C for 18 h, the surface area and pore size distribution of different samples were measured with the Brunauer-Emmett-Teller (BET) and Barrett-Joyner-Halenda (BJH) method utilizing the nitrogen adsorption-desorption analyzer (Quadrasorb EVO, Quantachrome). Vickers hardness and Young's modulus of samples were measured by indentation tests using nano indenter (Hysitron TI 980 Triboindenter, Bruker). Specifically, the change of the indentation depth was obtained by increasing the force to 10 mN at a loading rate of 2 mN s$^{-1}$ and a holding time of 5 s. Sonication test was performed by ultrasonic machine (Elmasonic S 120H, Elma) at a power of 1000 W for 10 min. Electrical conductivity was measured utilizing the van der pauw method (2002 Mutimeter, Keithley). Thermal gravimetric analysis (TGA, Q500, TA Instruments) was performed in an air atmosphere with a heating rate of 10 °C min$^{-1}$ ramping from room temperature to 900 °C.

**Electrochemical characterization**

To evaluate the electrochemical performance of Si-based anodes, CR2032 coin cells were fabricated in an argon-filled glove box. The negative electrodes were fabricated by slurry coating method on a copper current collector. The homogeneous slurry composed of the active materials, Super P, CMC/SBR at the mass ratio of 8:1:1, whose solid content was controlled at ~20 wt%. As a consequence, the mass loading of Si-based active materials (for SG, W@SG, W@SG-S80, W@SG-S40, SG-H40) was controlled at 1.5-1.7 mg cm$^{-2}$, corresponding to the electrode loading of 1.9-2.1 mg cm$^{-2}$. To further match industrial standards, W@SG electrodes with higher active mass loading (3.8-3.9 mg cm$^{-2}$) were also prepared. All of coin cells were then assembled with Li foil as counter and reference electrode, polypropylene (PP) as the separator, 1 M LiPF$_6$ in EC/DEC (1:1 vol%) and 10 wt% FEC as the electrolyte, and Si-based electrodes as the working electrode.

Electrochemical performance tests were carried out using a battery testing system (CT2001A, LAND) and an electrochemical station (VSP-300, Bio-Logic). In particular, galvanostatic charge/discharge (GCD) test was performed under LAND battery testing system with the potential window from 0.01 to 1.5 V (vs. Li/Li$^+$). The rate capability would be characterized at various current densities of 0.2, 0.5, 0.8, 1 and 2 A g$^{-1}$, after the completion of activation process (0.1 A g$^{-1}$, first 3 cycles). The mass specific capacities of all electrodes were calculated based on the amount of active mass loading. Cyclic voltammetry (CV) and electrochemical impedance spectroscopy (EIS) measurements were conducted on a VSP-300 electrochemical workstation. Specifically, CV tests were carried out at a scan rate of 0.1 mV s$^{-1}$ between 0.01 and 1.5 V, and EIS tests were performed in the frequency range between 100 kHz and 0.01 Hz.

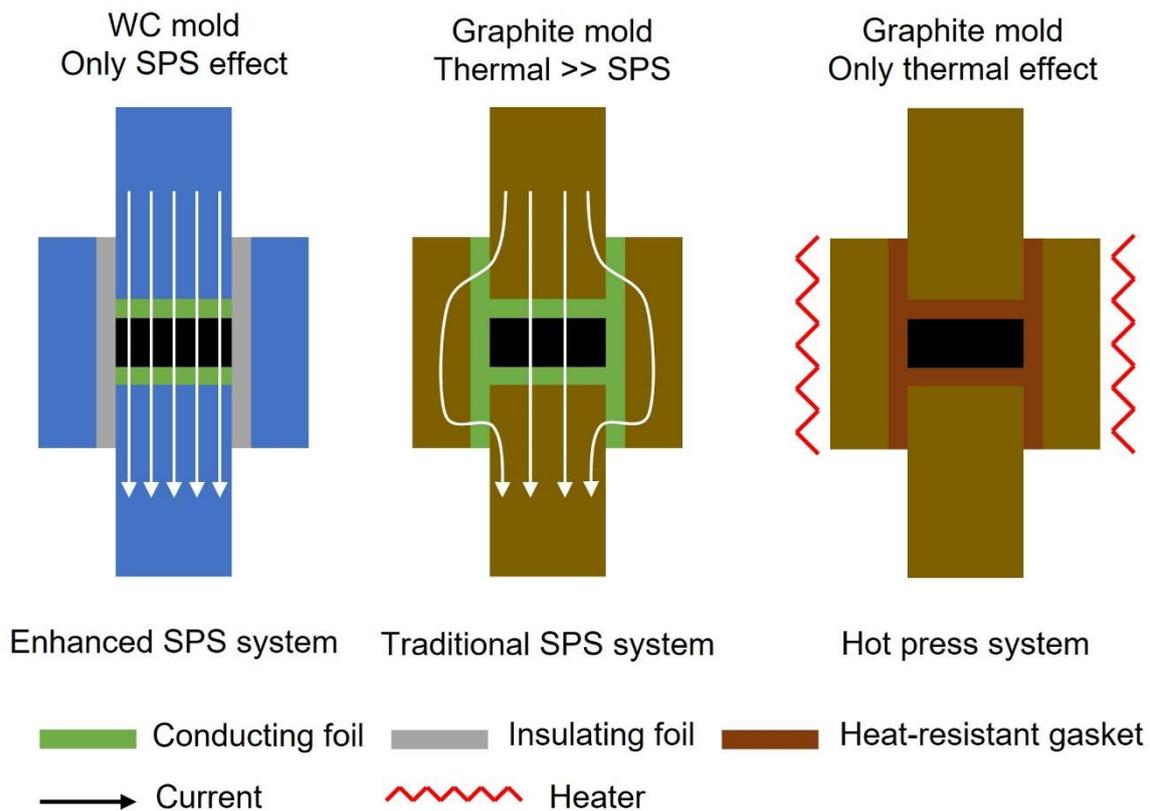

Figure S1. The working principle comparison diagram of the enhanced SPS system, traditional SPS system, and hot press system.

As shown in Figure S1, the enhanced SPS system utilizes a high electrical conductivity WC alloy as the working mold, assembled with an insulating foil on the side and conductive foils on the upper and lower sides to only allow the current flow through the powder during SPS operation, which only has the SPS effect. By contrast, the traditional SPS system employs a graphite mold with conductive foils surrounding the powder for protection. Since most of the current passes through the graphite mold during the SPS process (because the conductivity of the sample is less than that of the graphite mold), the thermal effect of the traditional SPS system is substantially bigger than the SPS effect. On the other hand, there is no electric current in the hot press system, whose sintering temperature is reached through the heater, thus only the thermal effect.

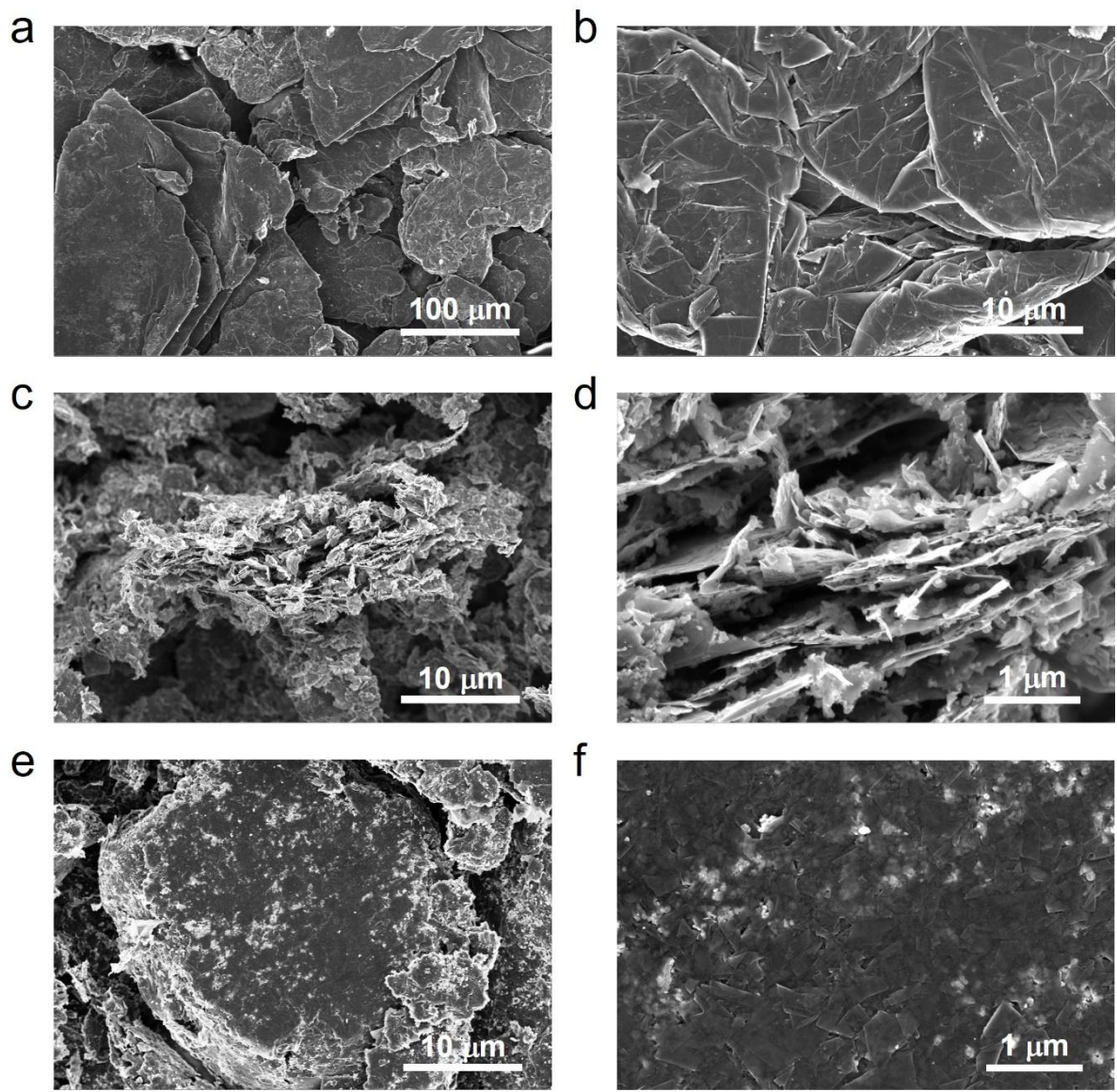

Figure S2. SEM images of (a, b) bare EG layers, (c, d) SG powder, and (e, f) W@SG blocks.

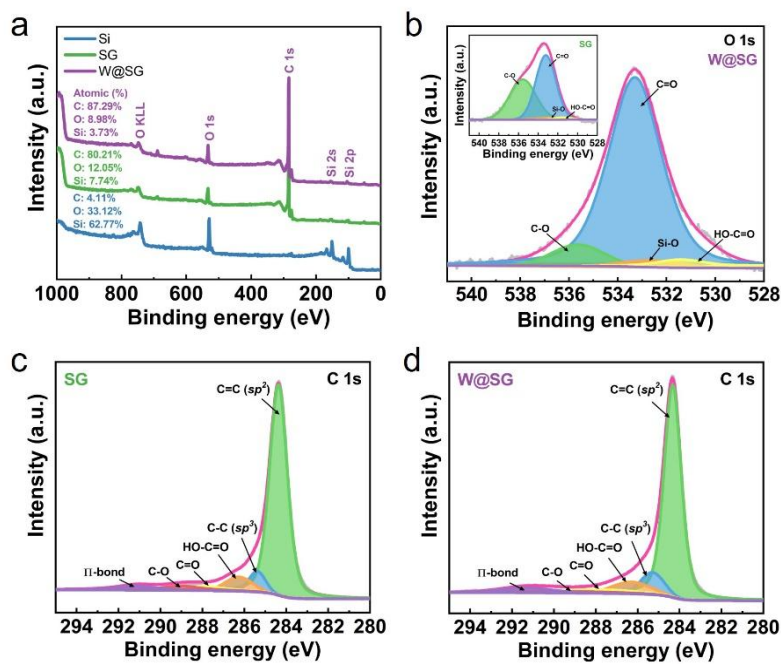

Figure S3. (a) XPS full spectrum of bare Si, SG, W@SG. (b) High-resolution XPS spectra of O 1s peaks of W@SG, inserted with that of SG. (c, d) High-resolution XPS spectra of C 1s peaks of SG and W@SG.

Based on the analysis of the full spectrum of XPS (Figure S3a), the atomic percentage of Si in SG is exceedingly low (7.74%) compared to the corresponding content in bare Si (62.77%). Since X-ray photoelectrons have a penetration depth of only a few nanometers, this can be compelling evidence for a carbon coating on the surface of SG. Furthermore, the atomic content of Si in W@SG is further reduced (3.73%) after enhanced SPS treatment, indicating that more compact carbon coverage on the surface, which can be attributed to the directional arrangement and distribution of EG sheets induced by enhanced SPS. More importantly, this compact covering structure is beneficial to improve the conductivity of the anode and avoid redundant side reactions. In addition, as can be seen from Figure S3b, the C-O peak of W@SG is significantly reduced compared with SG, indicating that enhanced SPS is beneficial to remove the functional groups of the material.[1] Furthermore, according to the peak fitting results of C 1s (Figure S3c and d), SG and W@SG have six characteristic signals corresponding to C=C ($sp^2$), C-C ($sp^3$), HO-C=O, C=O, C-O, and π-bond, respectively. The relative content of each peak in SG is 78.8%, 8.0%, 7.1%, 2.1%, 1.8%, and 2.2%, respectively; in W@SG it is 74.3%, 13.5%, 6.9%, 1.7%, 0.3%, and 3.3%, respectively. Notably, the relative content of C=C ($sp^2$) peak drops from 78.8% of SG to 74.3% of W@SG, while that of C-C ($sp^3$) increases from 8% to 13.5%, indicating an increase in structural disorder and interlayer bonding.[2] The reduction of the proportion of oxygen-containing functional groups (HO-C=O, C=O, and C-O) further confirmed the purification effect of enhanced SPS.

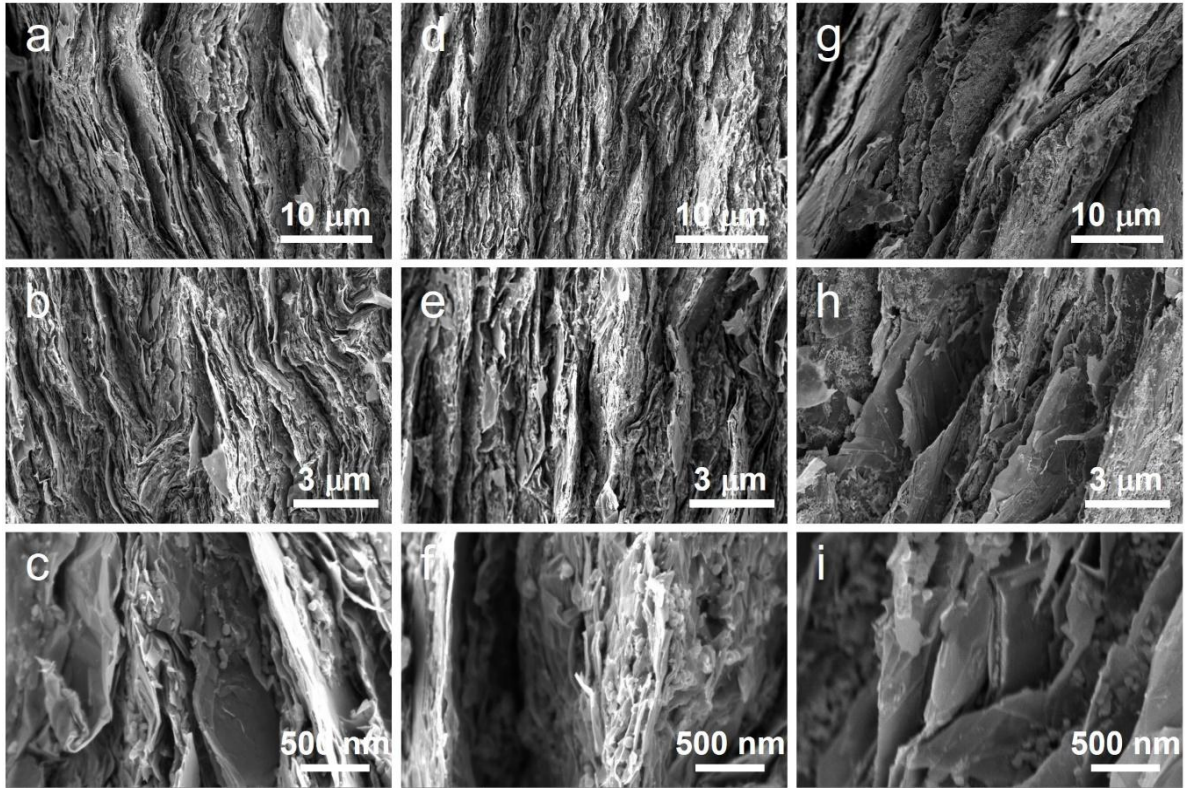

Figure S4. Cross-sectional SEM images of SG composite powder treated by different conditions. (a-c) W@SG-S80 (SPS, 800 °C, 80 MPa), (d-f) W@SG-S40 (SPS, 800 °C, 40 MPa), (g-i) SG-H40 (HP, 800 °C, 40 MPa).

Table S1. BET and BJH pore size analysis results of SG and W@SG

|  | **SG** | **W@SG** |
|---|---|---|
| Specific surface area (m$^2$ g$^{-1}$) | 53.7 | 37.8 |
| Total pore volume (cm$^3$ g$^{-1}$) | 0.274 | 0.078 |
| Average pore size (nm) | 20.4 | 8.3 |

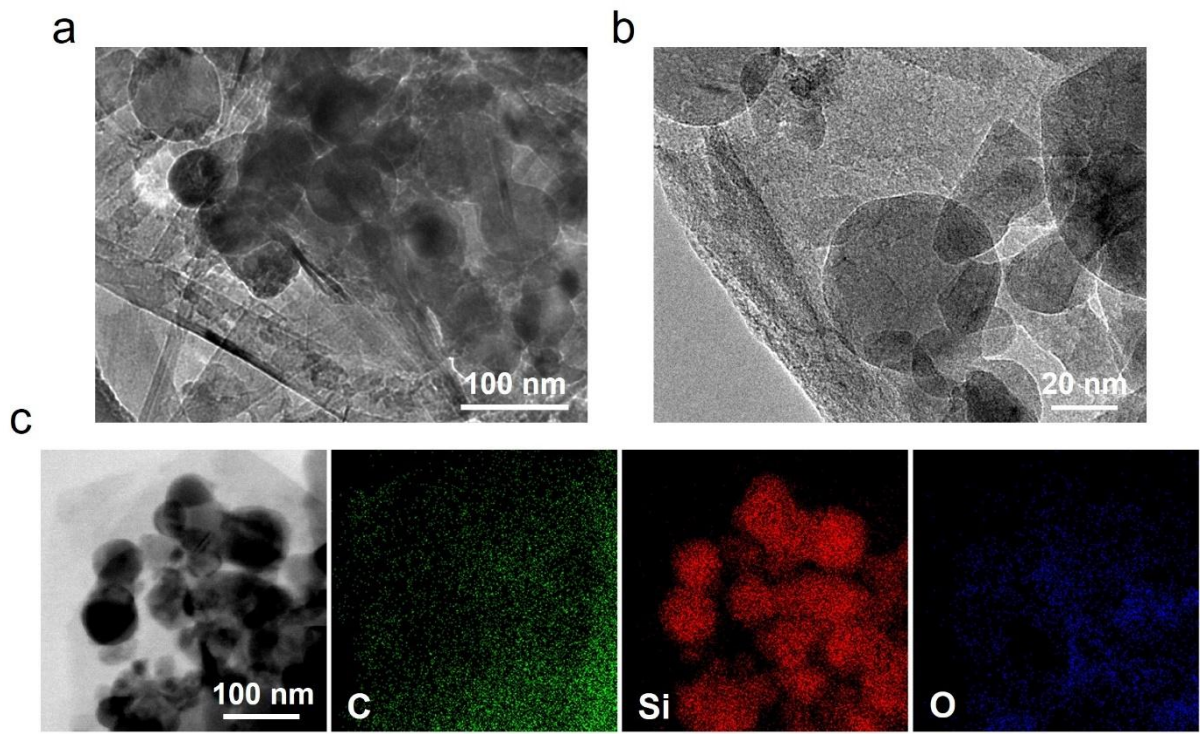

Figure S5. TEM images of SG with (a) low and (b) high magnification. (c) EDS mapping results of SG in the same area and relative intensities of C, Si, and O elements.

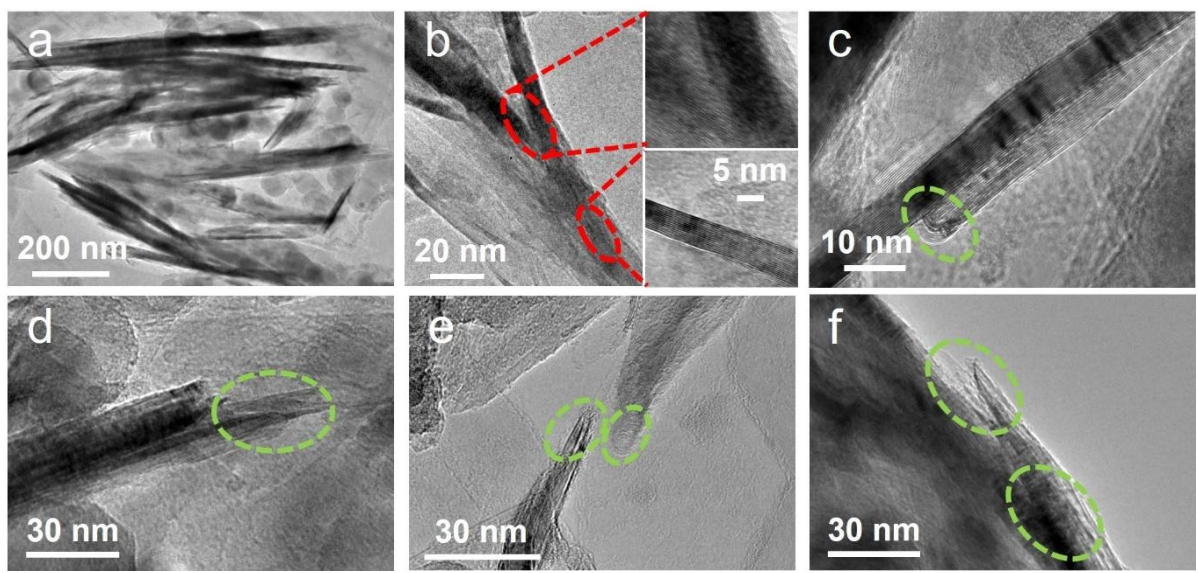

Figure S6. (a-f) TEM images of W@SG with different magnifications.

The black strips in Figure S6a are the welded structure of EG layers, since they are different from the normal horizontal lamination covering EG. Figure S6b depicts the evolution of the welding interface from two orientation at the beginning to one orientation, demonstrating that two EG layers are welded together. In addition, the green circles shown in Figure S6c-f indicate the closed-edge or rolled-up structure, resulting from interlayer-welding effect induced by enhanced SPS, which is similar to the structures induced by neutron and electron radiation reported in the literature.[3-7]

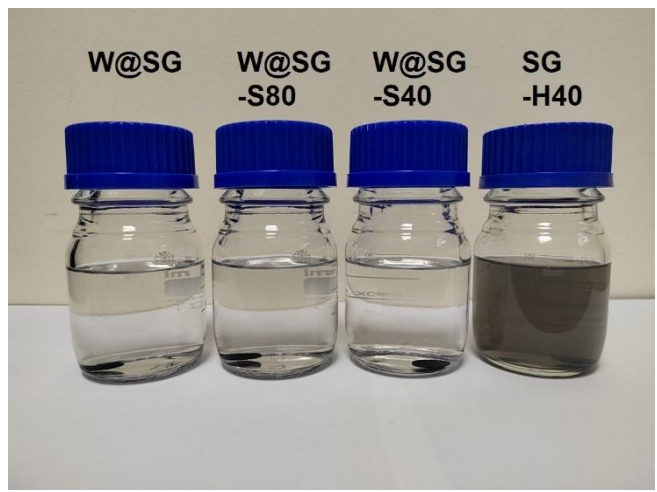

Figure S7. Sonication tests of W@SG, W@SG-S80, W@SG-S40, and SG-H40.

The as-prepared monolithic blocks after enhanced SPS or HP were performed in isopropanol solution by ultrasonic machine at a power of 1000 W for 10 min, to identify the mechanical stability. It can be seen from Figure S7 that W@SG, W@SG-S80, and W@SG-S40 blocks are not destroyed, and their solutions stay clear despite the intense ultrasonic shock, while that of SG-H40 becomes quite murky. This indicates that W@SG, W@SG-S80 and W@SG-S40 have more stable mechanical strength comparing with SG-H40.

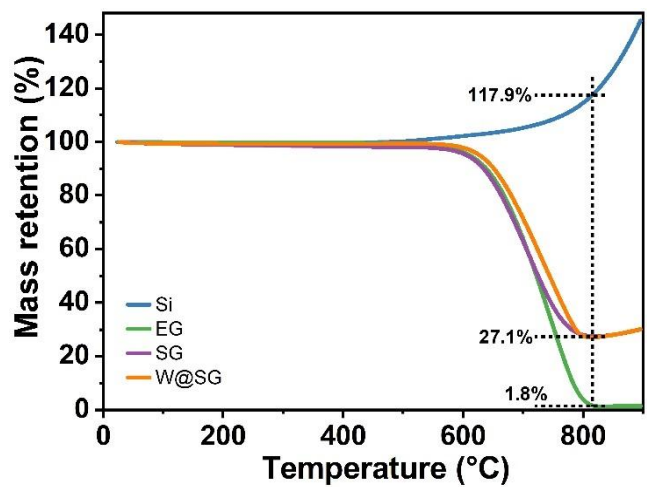

Figure S8. TGA curves of bare Si nanoparticles, EG, SG and W@SG samples.

As determined by the TGA results, EG is essentially burned out at 815 °C, leaving a residual of 1.8%, whereas pure Si is oxidized to 117.9% at the same temperature. With a value of 27.1%, the weight percentage of SG and W@SG reaches its lowest point, followed by a quick oxidation of Si to increase mass. Therefore, the mass ratio of Si in SG and W@SG calculated at 815 °C is 21.4%, which is close to the initial setting of the experiment (20%). Besides, the mass retention curve of SG decreases before that of W@SG, which can be taken as evidence that W@SG has higher thermal stability due to a more robust structure. Moreover, they exhibit the same lowest point, implying that their chemical compositions are consistent and enhanced SPS does not induce Si loss or oxidation.

Table S2. Comparison of Vickers Hardness and Young's modulus of W@SG with the reported Si-based anodes.

| Materials | Vickers Hardness (MPa) | Young's modulus (GPa) | Ref. |
|---|---|---|---|
| Conductive glue | 24.4 | 0.5 | [8] |
| Si@P-LiPAA | 25.6 | 1.0 | [9] |
| Si@N-P-LiPN | 15.2 | 0.9 | |
| Si@P-LiNF | 4.4 | 0.2 | |
| 3D-IBN@Si | 46.8 | 1.9 | [10] |
| Si@PVDF | 7.6 | 1.0 | [11] |
| Si@SHPET | 9.7 | 2.2 | |
| Si@CMC | 15.6 | 3.3 | |
| PFA-TPU/SiO$_x$ | 131 | 3.8 | [12] |
| PAA-GA@Si | 39.6 | 5.4 | [13] |
| TBA-VTLES10/Si | 132 | 5.9 | [14] |
| Si/Ti$_3$SiC$_2$ | 260 | 7.5 | [15] |
| W@SG | 658 | 11.6 | This work |

Table S3. Comparison of tap density of W@SG with the reported Si-based anodes.

| Materials | Tap density (g cm$^{-3}$) | Micro/Nano-sized Si | Journal/Year of publication | Ref. |
|---|---|---|---|---|
| Yolk-shell Si/C | 0.15 | Nano | Nat. Nanotechnol., 2014 | [16] |
| Si/Li$_2$Ti$_3$O$_7$/C | 0.35 | Nano | Energy Stor. Mater., 2021 | [17] |
| Blocky SiO$_x$/C | 0.69 | Micro | Adv. Funct. Mater., 2018 | [18] |
| Core-shell gradient porous Si | 0.70 | Micro | Energy Stor. Mater., 2021 | [19] |
| Si@SiO$_x$@C | 0.75 | Micro | Adv. Funct. Mater., 2021 | [20] |
| Watermelon-inspired Si/C | 0.88 | Nano | Adv. Energy Mater., 2017 | [21] |
| Diatomite-derived Si/SiO$_2$ | 0.90 | Micro | Adv. Funct. Mater., 2020 | [22] |
| Si@SiO$_2$@C | 0.91 | Nano | Energy Environ. Sci., 2015 | [23] |
| Nanoporous Si/C | 0.93 | Micro | Adv. Mater., 2021 | [24] |
| CNT-Si composite | 1.10 | Micro | Energy Environ. Sci., 2016 | [25] |
| Macropore-coordinated graphite-Si | 1.13 | Micro | Nat. Commun., 2019 | [26] |
| SiMg$_y$O$_x$@C | 1.15 | Micro | Adv. Mater., 2022 | [27] |
| W@SG | 1.68/1.12 (block/powder) | Nano | - | This work |

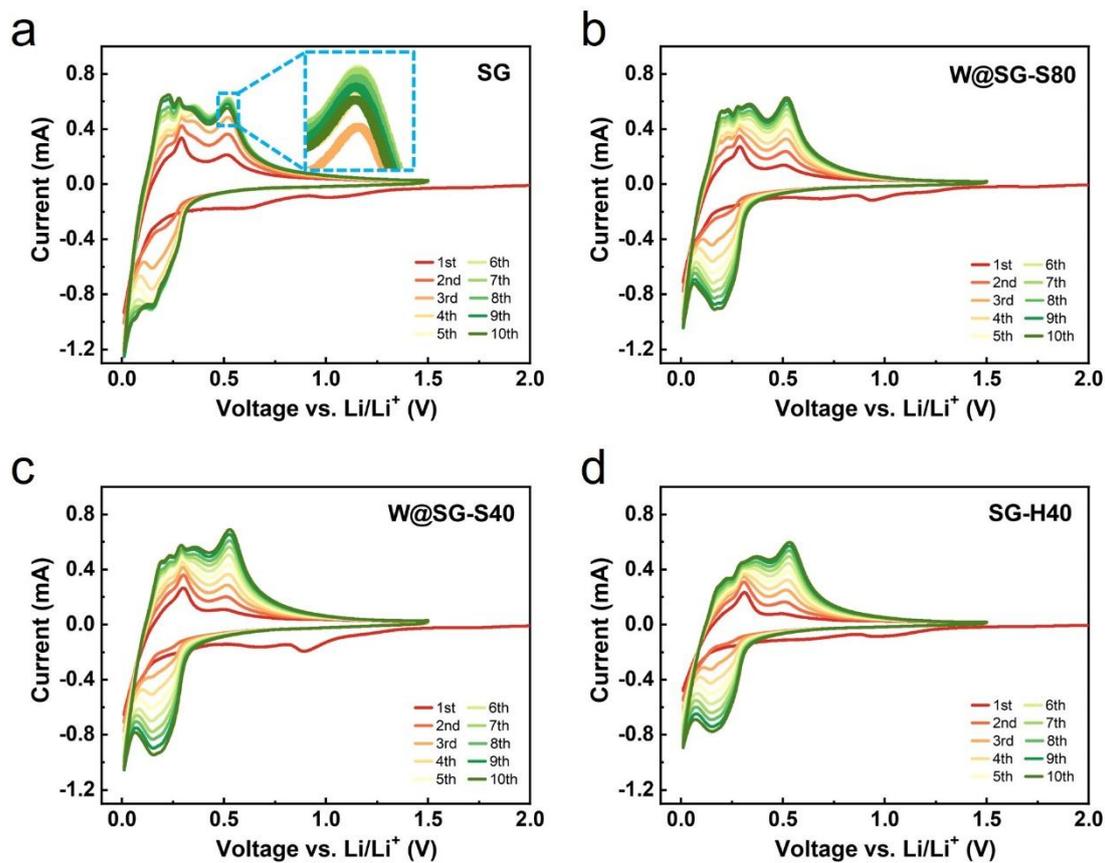

Figure S9. CV curves of (a) SG, (b) W@SG-S80, (c) W@SG-S40, and (d) SG-H40 electrodes at a scanning rate of 0.1 mV s$^{-1}$ with a potential window between 0.01 and 1.5 V (vs. Li/Li$^+$).

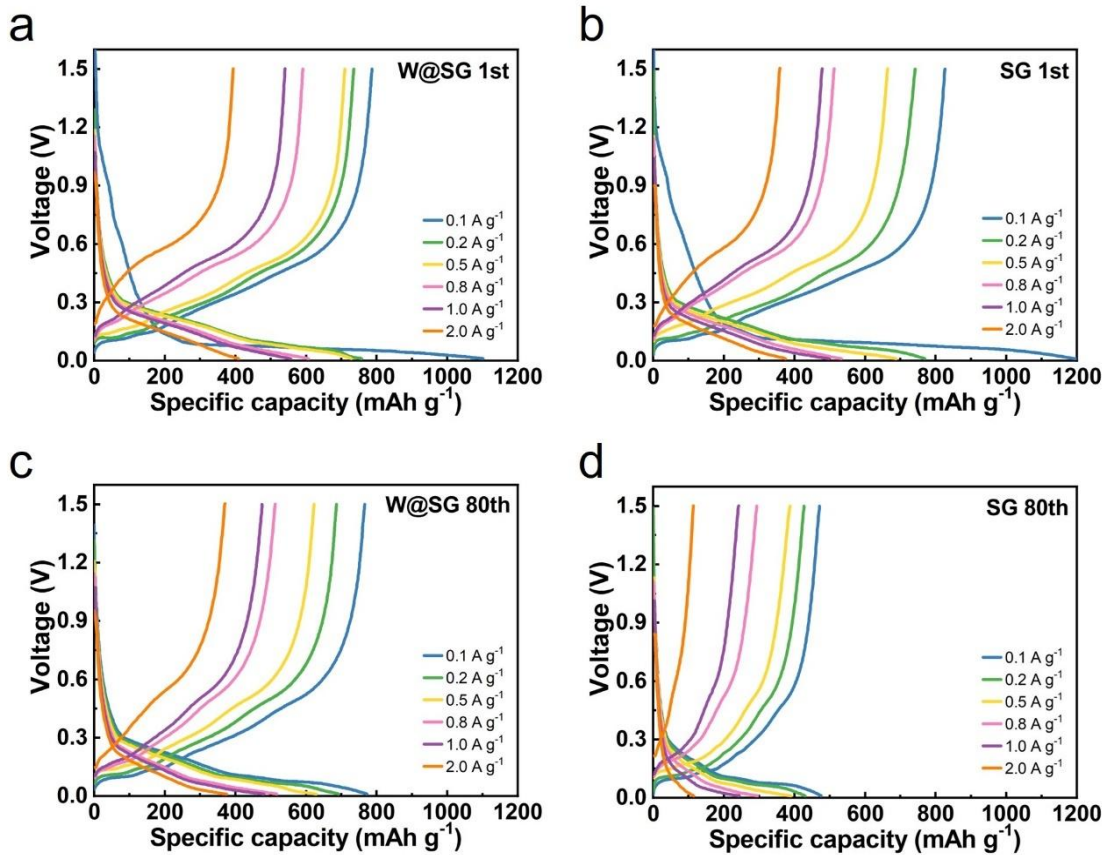

Figure S10. (a-d) Voltage profiles of W@SG (a, c) and SG (b, d) electrodes at the 1st and 80th cycles at various rates from 0.1 to 2 A g$^{-1}$ with a potential window between 0.01 and 1.5 V (vs. Li/Li$^+$).

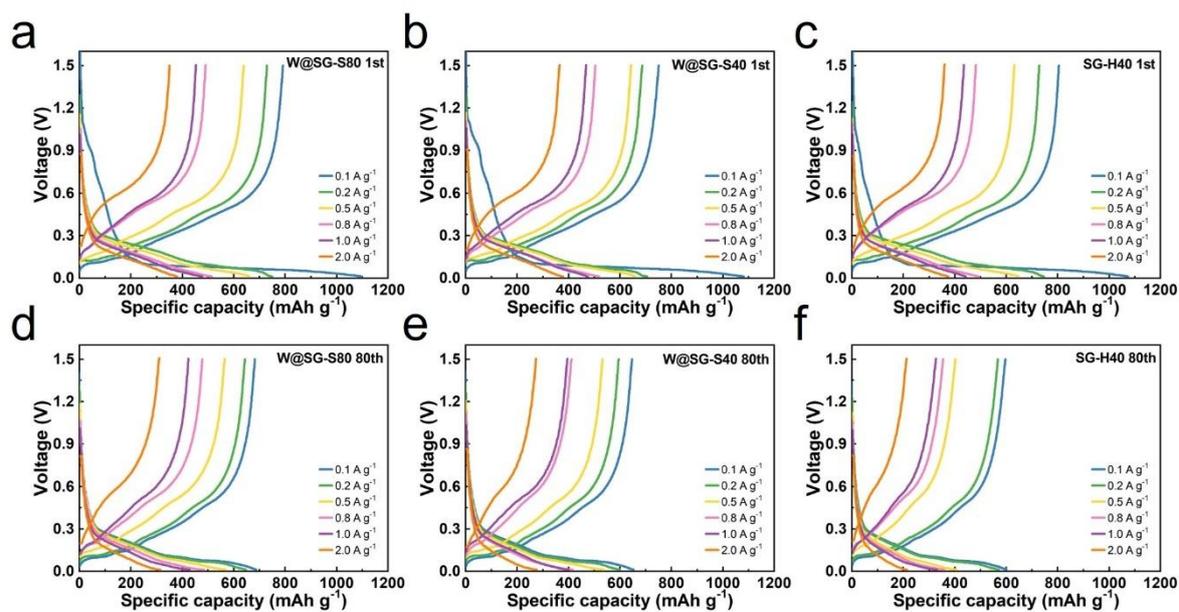

Figure S11. (a-f) Voltage profiles of W@SG-S80 (a, d), W@SG-S40 (b, e), and SG-H40 (c, f) electrodes at the 1st and 80th cycles at various rates from 0.1 to 2 A g$^{-1}$ with a potential window between 0.01 and 1.5 V (vs. Li/Li$^+$).

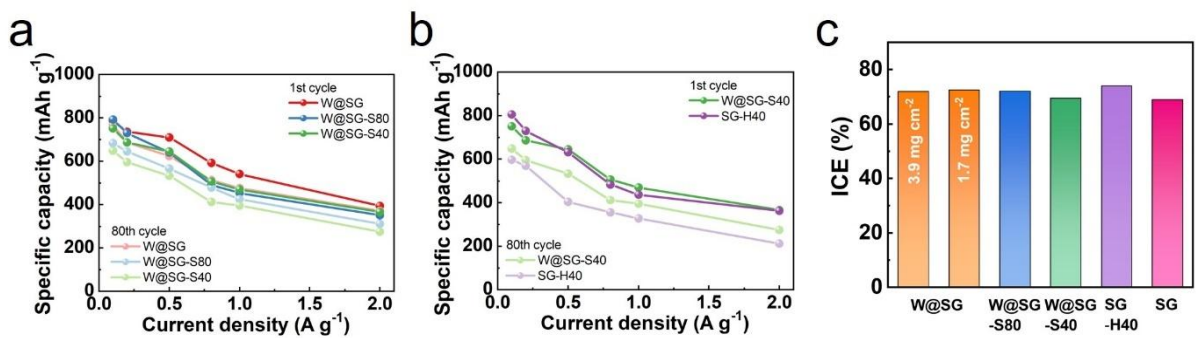

Figure S12. (a) The specific capacity variation curves of W@SG, W@SG-S80, and W@SG-S40 electrodes at the 1st and 80th cycles at different current densities. (b) The specific capacity variation curves of W@SG-S40 and SG-H40 electrodes at the 1st and 80th cycles at different current densities. (c) Comparison of the initial coulombic efficiency (ICE).

According to the voltage profiles (Figure S10 and S11), we compare the specific capacity changes of electrodes under different SPS conditions in Figure S12a, revealing that the SPS-treated samples all maintain excellent specific capacitance after 80 cycles at any rate. Among them, W@SG demonstrates the most outstanding rate and cycling performance. In particular, W@SG delivers a charging capacity of 786.9 mAh g$^{-1}$ at a low current density of 0.1 A g$^{-1}$, with an ultra-high retention of 97.4% after 80 cycles. Moreover, it still has a specific capacity of 393.6 mAh g$^{-1}$ at a high current density of 2 A g$^{-1}$, with a high retention of 94.2% after 80 cycles. The specific capacity variations of electrodes under enhanced SPS and traditional hot press conditions are shown in Figure S12b. Compared with the specific capacity at the 1st cycle, the decreasing range of the capacity in W@SG-S40 electrode is substantially less than that of SG-H40 at various rates, indicating that SPS effect brings higher electrochemical stability to the Si-based anode. Besides, Figure S12c shows the initial coulombic efficiency (ICE) of various Si-based anodes. It is worth noting that the ICE of the W@SG electrode remains almost constant (72.5%) even at high mass loading, which is also higher than that of the SG electrode (68.9%). The ICE of the anode is expected to be further improved when using micro-sized Si.

Table S4. The specific capacities and retention of the as-prepared Si-based electrodes under different current density and cycle number.

| Current density (A g$^{-1}$) | Cycle number | SG (mAh g$^{-1}$) | SG-H40 (mAh g$^{-1}$) | W@SG-S40 (mAh g$^{-1}$) | W@SG-S80 (mAh g$^{-1}$) | W@SG (mAh g$^{-1}$) |
|---|---|---|---|---|---|---|
| 2.0 | 1st | 358.3 | 361.3 | 365.0 | 351.0 | 393.6 |
| | 80th | 113.3 | 212.2 | 274.5 | 311.0 | 370.8 |
| | Retention (%) | 31.6 | 58.7 | 75.2 | 88.6 | 94.2 |
| 1.0 | 1st | 478.1 | 435.9 | 468.7 | 453.7 | 540.6 |
| | 80th | 241.9 | 327.0 | 395.5 | 424.5 | 475.8 |
| | Retention (%) | 50.6 | 75.0 | 84.4 | 93.6 | 88.0 |
| 0.8 | 1st | 512.4 | 482.0 | 505.0 | 491.3 | 591.0 |
| | 80th | 293.7 | 355.0 | 411.7 | 477.8 | 512.6 |
| | Retention (%) | 57.3 | 73.7 | 81.5 | 97.3 | 86.7 |
| 0.5 | 1st | 663.9 | 632.0 | 643.9 | 639.4 | 709.9 |
| | 80th | 387.0 | 403.2 | 532.7 | 564.8 | 622.9 |
| | Retention (%) | 58.3 | 63.8 | 82.7 | 88.3 | 87.7 |
| 0.2 | 1st | 741.6 | 728.7 | 686.5 | 729.6 | 735.3 |
| | 80th | 427.6 | 568.3 | 595.5 | 644.0 | 686.4 |
| | Retention (%) | 57.7 | 78.0 | 86.7 | 88.3 | 93.3 |
| 0.1 | 1st | 826.1 | 804.8 | 750.8 | 791.8 | 786.9 |
| | 80th | 471.4 | 597.3 | 647.3 | 682.8 | 766.2 |
| | Retention (%) | 57.1 | 74.2 | 86.2 | 86.2 | 97.4 |

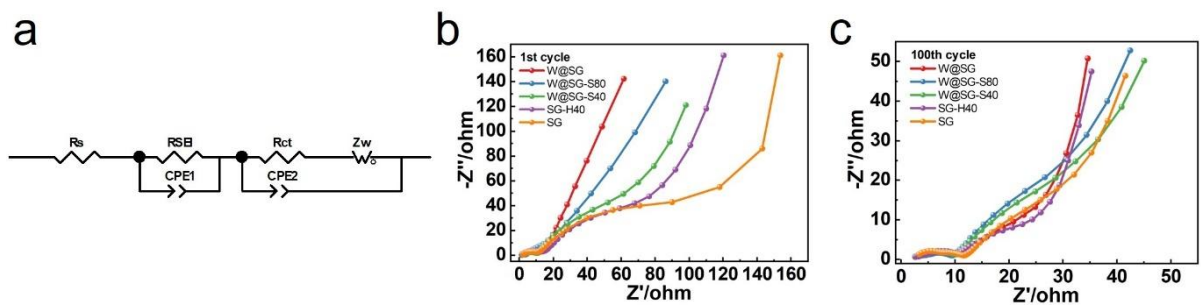

Figure S13. (a) Equivalent circuit model linked with EIS curves. EIS curves of the W@SG, W@SG-S80, W@SG-S40, SG-H40, SG electrodes after the 1st cycle (b) and 100th cycle (c).

Figure S13a shows the corresponding equivalent circuit model after EIS curve fitting of the as-prepared electrodes. In short, $R_s$ represents a set of resistances in the super-high frequency region that correspond to battery terminals, cables, and electrolyte resistance. The first semicircle at the high-frequency region reflects the SEI layer resistance on electrode ($R_{SEI}$), while the second semicircle in the middle-frequency region corresponds to charge transfer resistance ($R_{ct}$). The ion diffusion resistance ($R_{ion}$) is the Warburg impedance ($Z_w$) in the low frequency domain associated with Li$^+$ diffusion within the electrode, and CPE represents the interfacial constant phase element.

Table S5. The EIS simulation results of the as-prepared electrodes.

| | Cycle number | $R_s$ (Ω) | $R_{SEI}$ (Ω) | $R_{ct}$ (Ω) |
|---|---|---|---|---|
| **W@SG** | Fresh | 1.4 | 3.5 | 216.0 |
| | 1st | 1.4 | 2.1 | 6.1 |
| | 100th | 2.2 | 4.1 | 3.5 |
| **W@SG-S80** | Fresh | 1.5 | 4.8 | 303.5 |
| | 1st | 1.3 | 2.3 | 12.1 |
| | 100th | 2.2 | 8.9 | 4.1 |
| **W@SG-S40** | Fresh | 1.3 | 3.7 | 407.5 |
| | 1st | 1.8 | 10.2 | 84.2 |
| | 100th | 1.7 | 10.0 | 7.0 |
| **SG-H40** | Fresh | 1.1 | 5.6 | 552.6 |
| | 1st | 1.6 | 13.6 | 97.7 |
| | 100th | 1.8 | 10.7 | 8.3 |
| **SG** | Fresh | 1.2 | 10.1 | 604.1 |
| | 1st | 2.0 | 13.1 | 132.5 |
| | 100th | 3.4 | 12.0 | 15.0 |

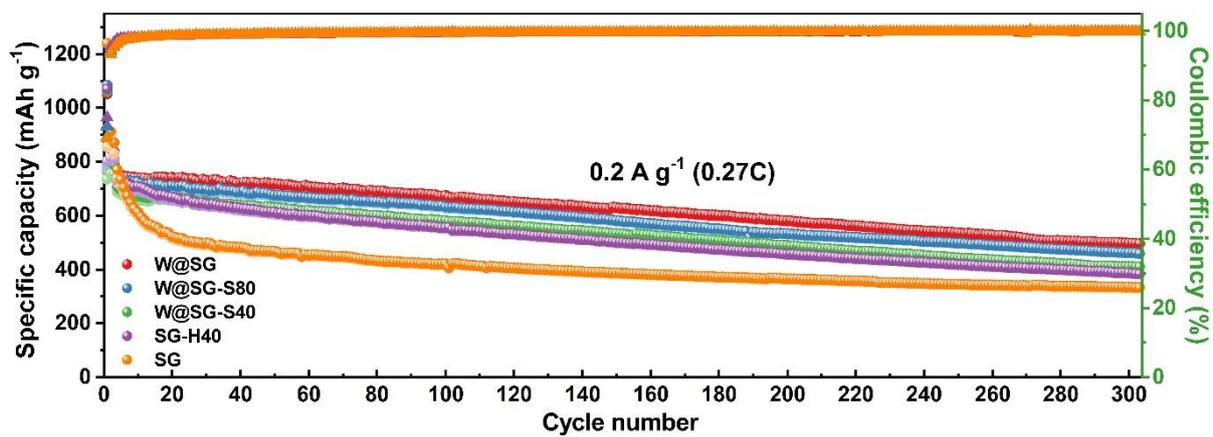

Figure S14. Long-term cycling performance of different electrodes (W@SG, W@SG-S80, W@SG-S40, SG-H40, and SG electrodes) at 0.2 A g$^{-1}$ (corresponding to 0.27C).

Table S6. Cycle life of batteries based on the as-prepared anodes.

| | SG | SG-H40 | W@SG-S40 | W@SG-S80 | W@SG-1.7 mg cm$^{-2}$ | W@SG-3.9 mg cm$^{-2}$ |
|---|---|---|---|---|---|---|
| Retention after 100 cycles at 0.1 A g$^{-1}$ (%) | 53.9 | 70.9 | 82.3 | 82.2 | 94.0 | 93.0 |
| Retention after 100/200 cycles at 2 A g$^{-1}$ (%) | 32.8/31.6 | 59.8/45.0 | 72.0/49.6 | 88.8/66.9 | 93.0/70.4 | - |

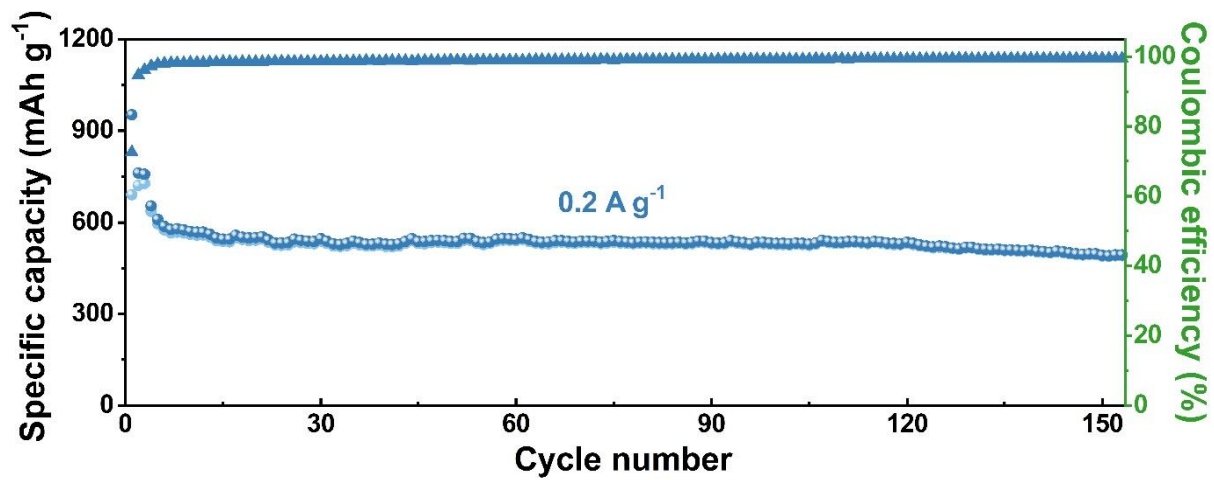

Figure S15. Long-term cycling performance of the W@SG-3.9 mg cm$^{-2}$ electrode at 0.2 A g$^{-1}$.

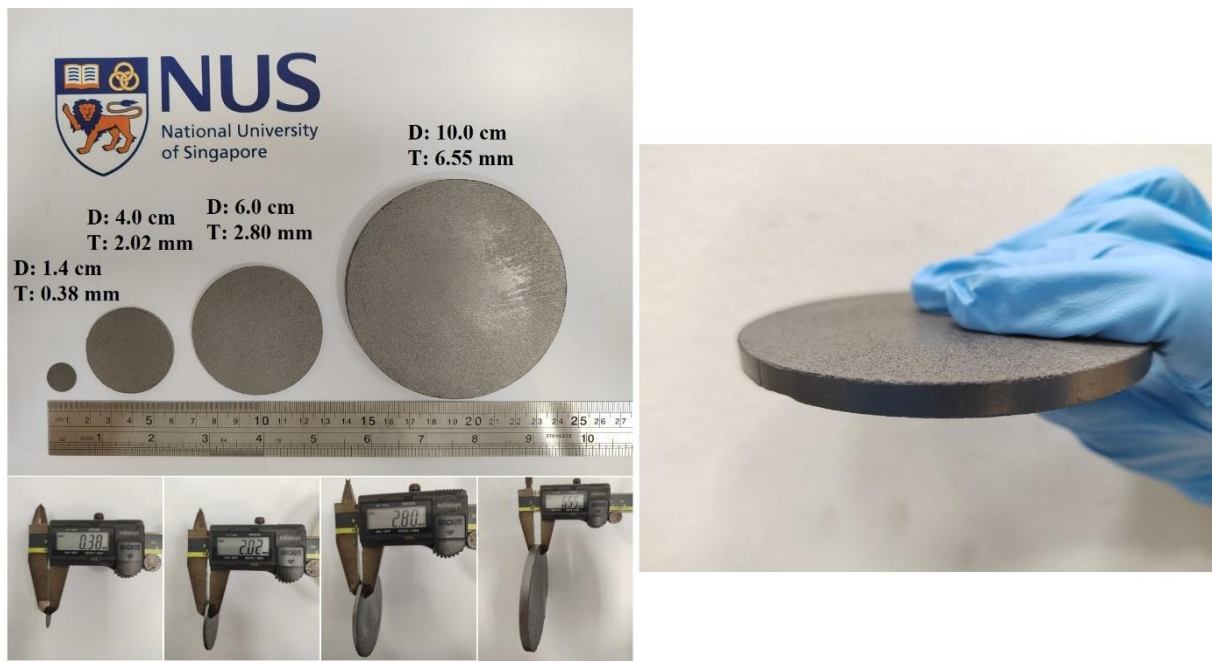

Figure S16. Photographic images of the monolithic W@SG blocks with different thicknesses and sizes.

Table S7. Comparison of loading level, areal capacity, and cycle life of the W@SG anode with the reported Si-based anodes.

| Si-based materials | Active mass loading (mg cm$^{-2}$) | Areal capacity (mAh cm$^{-2}$) | Current density | Retention (%) after 100th cycle | Ref. |
|---|---|---|---|---|---|
| Granadilla-like porous-structured Si/C | 1.3 | 1.2 | 0.1 A g$^{-1}$ | 69 | [28] |
| Mg$_2$Si-inspired nanoporous Si | 1.4 | 2.8 | 0.2 A g$^{-1}$ | ~69 | [29] |
| Si@CNT/C-microscroll | 2.2 | 5.6 | 0.2 A g$^{-1}$ | 78 | [30] |
| Si@SiO$_x$@C | 1.2 | 2.6 | 1.0 A g$^{-1}$ | 80 | [20] |
| Si/CNT | 1.0 | 1.6 | 0.2 A g$^{-1}$ | 81 | [31] |
| B-Si/CNT@G | 11.2 | 5.2 | 0.1 A g$^{-1}$ | 83 | [32] |
| Core-shell gradient porous Si | 0.8 | 1.7 | 1.0 A g$^{-1}$ | 84 | [19] |
| Si@C/EG | 1.2 | 1.1 | 0.2 A g$^{-1}$ | 85 | [33] |
| Blocky SiO$_x$/C | 3.5 | 2.3 | 0.5 C | ~87 | [18] |
| Watermelon-inspired Si/C | 4.1 | 2.5 | 0.2 A g$^{-1}$ | ~88 | [21] |
| Si/G/C microparticles | 2.0 | 1.2 | 0.2 C | 89 | [34] |
| Graphene-scaffolded Si/graphite | 1.1 | 0.6 | 0.4 A g$^{-1}$ | ~90 | [35] |
| Yolk-shell Si/C | 2.8 | 2.5 | 0.2 A g$^{-1}$ | 90 | [36] |
| CNT-Si | 1.3 | 2.8 | 0.4 A g$^{-1}$ | 90 | [25] |
| Branched Si@C | 0.8 | 1.9 | 0.02 C | 92 | [37] |
| Dual-shell SiNPs@C | 3.1 | 1.9 | 0.25 A g$^{-1}$ | ~93 | [38] |
| Si-graphite | 6.2 | 2.5 | 0.3 A g$^{-1}$ | ~94 | [39] |
| Nonfilling carbon-coated porous Si | 2.0 | 2.8 | 0.12 A g$^{-1}$ | 94 | [40] |
| Pomegranate-inspired Si | 3.1 | 3.0 | 0.22 A g$^{-1}$ | 94 | [16] |
| W@SG | 3.9 (Active loading) 4.8 (Electrode loading) | 2.9 | 0.1 A g$^{-1}$ | 93 | This work |